\documentclass[nofootinbib, reprint, amsmath,amssymb, aps]{revtex4-2}
\usepackage{graphicx}
\usepackage{dcolumn}
\usepackage{bm}
\usepackage[colorlinks]{hyperref}

\begin{document}


\title{Statistical Analysis of Stochastic Magnetic Fields}

\author{Amir Jafari}
 \email{elenceq@jhu.edu}
\author{Ethan Vishniac}%
 \email{evishni1@jhu.edu}
 
 \author{Vignesh Vaikundaraman }%
 \email{vvaikun1@jhu.edu}

\affiliation{%
Johns Hopkins University, Baltimore, USA\\
}%

\date{\today}

\begin{abstract}

Previous work has introduced scale-split energy density $\psi_{l,L}({\bf x}, t)={1\over 2}{\bf{B}}_l.{\bf{B}}_L$ for vector field ${\bf B}({\bf x}, t)$ coarse-grained at scales $l$ and $L$, in order to quantify the field stochasticity. In this formalism, the ${\cal L}_p$-norms $S_p(t)={1\over 2}|| 1-\hat{\bf B}_l.\hat{\bf B}_L||_p$, $p$th order stochasticity level, and  $E_p(t)={1\over 2}|| B_l B_L||_p$, $p$th order mean cross energy density, are used to analyze the evolution of the stochastic field ${\bf B}({\bf x}, t)$. Application to turbulent magnetic fields leads to the prediction that turbulence in general tends to tangle an initially smooth magnetic field increasing the magnetic stochasticity level, $\partial_tS_p\geq 0$. An increasing magnetic stochasticity in turn leads to disalignments of the coarse-grained fields ${\bf B}_d$ at smaller scales, $d\ll L$, thus they average to weaker fields ${\bf B}_L$ at larger scales upon coarse-graining, i.e., $\partial_t E_p\leq 0$. Magnetic field resists the tangling effect of the turbulence by means of magnetic tension force. This can lead at some point to a sudden slippage between the field and fluid, decreasing the stochasticity $\partial_tS_p\leq 0$ and increasing the mean cross energy density $\partial_tE_p\geq 0$. Thus the maxima (minima) of magnetic stochasticity are expected to approximately coincide with the minima (maxima) of cross energy density, occurrence of which corresponds to slippage of the magnetic field through the fluid. Field-fluid slippage, on the other hand, has already been found to be intimately related to magnetic reconnection. Hence in this formalism, magnetic reconnection/slippage corresponds to $T_p=\partial_t S_p=0 \;\&\; \partial_t T_2<0$. In this paper, we test these theoretical predictions numerically using a homogeneous, incompressible magnetohydrodynamic (MHD) simulation. Apart from expected small scale deviations, possibly due to e.g., intermittency and strong field annihilation, the theoretically predicted global relationship between stochasticity and cross energy is observed in different sub-volumes of the simulation box. This may indicate ubiquitous local field-fluid slippage and reconnection events in MHD turbulence. We also show that the conditions $T_p=\partial_t S_p=0 \;\&\; \partial_t T_p<0$ lead to sudden increases in kinetic stochasticity level, i.e, $\tau_p=\partial_t s_p(t)>0$ with $s_p(t)={1\over 2}|| 1-\hat{\bf u}_l.\hat{\bf u}_L||_p$, which may correspond to fluid jets spontaneously driven by sudden field-fluid slippages---magnetic reconnection. This picture suggests defining reconnection as field-fluid slippage (changes in $S_p$) accompanied with magnetic energy dissipation (changes in $E_p$). All in all, these provide a new statistical approach to the reconnection in terms of the time evolution of magnetic and kinetic stochasticities, $S_p$ and $s_p$, their time derivatives, $T_p=\partial_t S_p$, $\tau_p=\partial_t s_p$, and corresponding cross energies, $E_p$, $e_p(t)={1\over 2}|| u_l u_L||_p$. Furthermore, we introduce the scale-split magnetic helicity, based on which, we discuss the stochasticity relaxation of turbulent magnetic fields. Finally, we construct and numerically test a toy model, which resembles a classical version of quantum mean field Ising model for magnetized fluids, in order to illustrate how turbulent energy can affect magnetic stochasticity in the weak field regime.
\end{abstract}

\pacs{Valid PACS appear here}
\maketitle


\section{\label{sec:level1}Introduction}

Perhaps the most famous example of unstable solutions for differential equations in physics is Einstein's static solution to the field equations in general relativity. Einstein had to introduce an extra term, proportional to the metric and a constant called the cosmological constant, in order to achieve a static solution describing the whole universe (this was in the early $1900s$ and the expansion of universe had not been discovered yet). His static solution turned out later to be unstable like a pen balanced on its tip; a small perturbation would lead to either an expanding or a contracting universe \footnote{Observations by Slipher, and later on by Hubble, revealed redshift in the spectrum of cosmologically distant objects implying that the universe was in fact expanding and not static. Hence there was no need to introduce the cosmological constant at all, which was why Einstein regretted adding it to his field equations. But there was a more serious problem. Quantum fluctuations in vacuum give empty space energy; the sum of all ground states of quantum fields. This energy should gravitate according to general relativity. Therefore, the effective cosmological constant $\lambda_{eff}$ is the sum of all such zero-point energies of normal modes of quantum fields $\langle \rho_{QFT}\rangle $ and Einstein's constant $\lambda$. According to cosmological observations, these two constants cancel out to better than $118$ decimal places leading to a very small $\lambda_{eff}=\lambda+\langle \rho_{QFT}\rangle $. This fine-tuning problem is the modern cosmological constant problem; see \citep*{Weinberg1989} for a classical review.}. Similar situations arise in other problems, for example, as viscosity tends to zero in a fluid, i.e., the Reynolds number tends to infinity, the hydrodynamic solutions become unstable. Physically, this translates into the fact that as viscosity becomes smaller and smaller, or the Reynolds number larger and larger, the flow becomes more sensitive to the development of turbulence \footnote{The simple mathematical fact that $\nu \rightarrow 0$ is totally different from $\nu=0$, despite its triviality, is sometimes overlooked raising confusion and misunderstanding. In asserting that in the limit of small viscosity $\nu\rightarrow 0$, or large Reynolds number $Re\rightarrow \infty$, the flow becomes unstable to the development of turbulence, there is no need for zero viscosity which is physically implausible of course. Instead, what is implied is that one can take the viscosity as small as, or the Reynolds number as large as, one wishes.}. For instance, a cup of coffee (small viscosity) can easily become turbulent if stirred by a spoon while a cup of honey, with a much larger viscosity than coffee, will retain its laminar flow even if stirred forcefully.  However, large amounts of honey (large Reynolds numbers) can still become turbulent although its viscosity is much larger than that of coffee. This also explains why lava, which is typically $10^5$ times more viscous than water, can become turbulent as it flows out of an active volcano. Even if viscosity $\nu$ is not small at all, therefore, a system with a large length scale $L$ or a large characteristic velocity $U$ can become turbulent since the corresponding Reynolds number $Re=LU/\nu$ can be huge anyway. Mathematically, any MHD equation, e.g., Navier-Stokes or the induction equations, can be written in terms of some dimensionless variables such as

$${\overline {\bf x}}={\bf x}/L,\;\;\;{\overline t}=t/(L/U),\;\;\;{\overline{\bf u}}={\bf u}/U,\;\;\;{\overline p}=p/U^2,$$

with pressure $p$. The Navier-Stokes equation, for instance, becomes

$${\partial \overline {\bf u}\over \partial \overline t}+\overline{\bf u}.\overline\nabla\overline{\bf u}=-\overline\nabla\overline p+{1\over Re}\overline\nabla^2\overline{\bf u}.$$

Therefore, the regime of small viscosity $\nu$ (resistivity $\eta$) translates into the regime of large kinetic (magnetic) Reynolds number $Re=UL/\nu$ ($Re_m=UL/\eta$). That is to say, the limit $\nu\rightarrow 0$ ($\eta\rightarrow 0$) is equivalent to the limit $Re\rightarrow \infty$ ($Re_m\rightarrow \infty$). In astrophysical systems, $L$ is usually very large thus both kinetic and magnetic Reynolds numbers are huge, e.g., of order $10^{14}$ in a galaxy. This implies that such systems are typically turbulent.

The rate at which energy is dissipated by viscosity is given by $\nu|\nabla{\bf u}|^2$, thus we might expect that in the limit when viscosity becomes negligibly small, i.e., $\nu\rightarrow 0$ or $Re\rightarrow \infty$, the dissipation rate would vanish. Interestingly, it does not! This means that the velocity gradients should have blown up $|\nabla{\bf u}|\rightarrow \infty$ so in the limit $\nu\rightarrow 0$, the dissipation rate remains finite $\lim_{\nu\rightarrow 0}\nu|\nabla{\bf u}|^2\rightarrow \epsilon>0$ (see \cite{Eyink2018} and references therein). Because the limit $\nu\rightarrow 0$, or $Re\rightarrow \infty$, usually corresponds to turbulent motions, thus we conclude that in turbulence, the velocity field would in general become H{\"o}lder singular\footnote{A real valued function $g$ in ${\cal{R}}^n$ is H{\"o}lder continuous if two non-negative and real constants $C$ and $h$ exist such that $| g(x) - g(y) | \leq C\| x - y\|^h$ for all $x, y \in Domain(g)$. If the H{\"o}lder exponent $h$ is equal to unity, then $g$ is Lipschitz continuous. If $h< 1$, $g$ is called H{\"o}lder singular.} and its gradients would blow up $|\nabla{\bf u}|>\infty$. It turns out that in fact particle (Lagrangian) trajectories become stochastic (random; indeterministic and non-unique) under these conditions (see e.g., \citep{Bernardetal.1998}; \citep*{GawedzkiandVergassola2000}; \citep*{VandenEijndenandVandenEijnden2000}). This means that similar to the uncertainty encountered in quantum mechanics, one cannot predict the exact trajectory of any fluid particle with certainty. This is a remarkable fact by itself (since it implies that God plays dice even in classical physics), but it also has extremely important consequences for magnetic fields. 

Usually, in particular in astrophysical setups, large kinetic Reynolds numbers are  accompanied with large magnetic Reynolds numbers. The ratio of the latter to the former is, in fact, the ratio of viscosity to resistivity---magnetic Prandtl number $Pr_m=\nu/\eta$. Analogous to velocity field in the limit of vanishing viscosity, magnetic field as well in the limit of vanishing resistivity $\eta\rightarrow$, i.e., $Re_m\rightarrow \infty$, becomes H{\"o}lder singular as magnetic energy dissipation rate does not vanish either $\eta|\nabla {\bf B}|^2\rightarrow \epsilon_m>0$, as experiments and simulations indicate \cite{JV2019} and references therein. Magnetic field gradients become ill-defined consequently; $|\nabla{\bf B}|\rightarrow \infty$ when $\eta\rightarrow 0$. The Alfv\'en flux freezing theorem \citep*{Alfven1942} indicates that, in the limit of vanishing resistivity, $\eta\rightarrow 0$, magnetic field lines will be frozen into the fluid. However, in a turbulent fluid, the velocity field is H{\"o}lder singular and the corresponding particle trajectories are indeterministic and non-unique: which particle trajectory will magnetic field lines follow? Magnetic fields will behave stochastically under such conditions (\citep*{Eyink2011}; \citep*{Eyinketal.2013}; \citep*{Eyink2015}; \citep*{Lalescuetal.2015}) and consequently magnetic field lines will not have any identity preserved in time, i.e., it is impossible to pick up a field line and track it in time. Even the definition of magnetic field lines as parametric curves ${\bf x}(s)$ whose tangent lines are given by the magnetic field ${\bf B}({\bf x}, t)$ at any point $({\bf x}, t)$ breaks down since the corresponding differential equation $d{\bf x}(s)/ds=\hat{\bf B}({\bf x}(s), t)$ has non-unique solutions for H{\"o}lder singular $\hat{\bf B}={\bf B}/|\bf B|$ \citep*{JV2019}. In such situations, the conventional flux freezing would not apply, instead a stochastic version of the Alfv\'en theorem, recently developed by Eyink \citep*{Eyink2011}, is required.

The above arguments have been established rigorously and made mathematically precise over the last few decades (see e.g. \cite{JV2019}; \cite{JVV2019}; \cite{Eyink2018} and references therein). In the presence of turbulence, it is now well understood that magnetic field behaves stochastically and its evolution should be studied in a statistical sense. The key points to keep in mind in particular include the fact that magnetic field is not frozen into the fluid in the conventional sense (rather it holds in a statistical sense); there is field-fluid slippage on a wide range of scales and the fact that magnetic reconnection---magnetically driven spontaneous, eruptive fluid motions---occurs not only in small diffusion regions but also on much larger scales in the turbulence inertial range see \S\ref{sreconnection}). Finally, since neither magnetic field nor velocity field is (Lipschitz) continuous in turbulent flows, as discussed above, thus their spatial derivatives may be ill-defined. These (H{\"o}older) singularities in MHD equations, which contain all sorts of such spatial derivatives, can be removed by, for example, coarse-graining, which will be briefly reviewed in \S\ref{sMagnetic Stochasticity}.

Stochastic flux freezing \citep*{Eyink2011} (generalized version of Alfv\'en flux freezing applicable also in turbulence) along with the notion of the stochasticity of field lines (\citep*{EyinkandAluie2006};  \citep*{Eyinketal.2013}; \citep*{Lalescuetal.2015}) play crucial roles in the evolution of magnetic fields including the phenomenon of magnetic reconnection. In recent years, the problem of magnetic reconnection (for a review of magnetic reconnection see e.g., \citep*{DorelliandBhattacharjee2008}; \citep*{Yamadaetal.2016}; \citep*{LoureiroandUzdensky2016}; \citep*{JafariandVishniac2018}) in turbulence has been approached taking into account the field stochasticity (see e.g., \citep*{LazarianandVishniac1999}; \citep*{Jafarietal.2018}). Yet concepts such as topology change and weak/strong stochasticity are widely used without providing concise mathematical definitions. Jafari and Vishniac \citep*{JV2019} have recently provided a rigorous mathematical definition for magnetic stochasticity, or spatial complexity, in terms of the renormalized, i.e., coarse-grained, magnetic field at different scales.  Magnetic field ${\bf B}({\bf x}, t)$ is coarse-grained, or renormalized, at scale $l$ by multiplying it by a rapidly decaying function (kernel) $G({\bf r})$ and integrating: ${\bf B}_l({\bf x}, t) =\int_V G_l({\bf r}){\bf B}({\bf x + r}, t)d^3r$. This is the average magnetic field of a parcel of fluid of length-scale $l$ at point $({\bf x}, t)$. In general, ${\bf B}_l({\bf x}, t)$ will differ from ${\bf B}_L({\bf x}, t)$ for $l\neq L$. For a stochastic field ${\bf B}({\bf x}, t)$ (in turbulence), the angle between ${\bf B}_l({\bf x}, t)$ and ${\bf B}_L({\bf x}, t)$ at any arbitrary point ${\bf x}$ will fluctuate as a stochastic variable. Therefore, $\phi({\bf x}, t) = \cos\theta = \hat{\bf B}_l.\hat{\bf B}_L$ is a measure of local magnetic stochasticity at point $({\bf x},t)$. The rms-average of $(1-\phi)/2$ is a time-dependent, volume-averaged function which measures magnetic stochasticity level in a volume $V$: $S(t) = (1-\phi)_{rms}/2$. The temporal changes in the stochasticity level in turn define topological deformations of the magnetic field and can be related to magnetic topology change. A short review of these concepts is given in \S\ref{rev}. A quantitative relationship between magnetic stochasticity or spatial complexity $S(t)$ and magnetic diffusion in turbulence has also established in \cite{JVV2019}. In this paper, we extend and illustrate this mathematical formalism using physical arguments and a toy model, to show how the topology and energy content of a turbulent magnetic field are related to its stochasticity level. In particular, we use an incompressible, homogenous MHD simulation, archived in an online, web-accessible database (\citep*{JHTDB};\citep*{JHTB1};\citep*{JHTB2}), to test the predictions of this model for magnetic reconnection and the slippage of magnetic field through the fluid.

We should emphasize that this paper presents only a statistical, rather than a diagnostic, analysis based on the statistical behavior of turbulent magnetic fields. Thus our approach does not involve the dynamics governed or affected by different instabilities, which in fact may accompany or even trigger magnetic reconnection. It is thus beyond the scope of this paper to study such phenomena including, in particular, tearing modes instability which, incidentally, turns out to play a subdominant role in 3D reconnection as already shown by many authors (e.g., see \cite{LazarianandVishniac1999}; \cite{Eyinketal.2011}; \cite{Eyinketal.2013} and references therein).

The detailed plan of the present paper is as follows: In  \S\ref{sMagnetic Stochasticity}, we review the method of coarse-graining used to remove magnetic field singularities and the theoretical approach to formulate stochasticity level of magnetic fields developed by Jafari and Vishniac \citep*{JV2019}. Also, a brief introduction to magnetic field-fluid slippage \citep*{Eyink2015} and magnetic reconnection with a focus on stochastic reconnection \citep*{LazarianandVishniac1999} is provided in this section. In \S\ref{sStatistical Study of Stochasticity}, we consider the field-fluid slippage and reconnection in MHD turbulence and extend the results of \citep*{JV2019} by (a) defining the kinetic stochasticity $s_p$ and cross energy $e_p$ and relating the time evolution of them to their magnetic counterparts, (b) constructing a toy model for weak magnetic fields in analogy with the classic mean field Ising model for magnetic spins and finally (c) defining the scale-split magnetic helicity and applying it to turbulent fields to study magnetic stochasticity and energy relaxation. These theoretical results are then tested using an incompressible, homogeneous MHD simulation stored online in Johns Hopkins Turbulence Database (\citep*{JHTDB};\citep*{JHTB1};\citep*{JHTB2}). We summarize and discuss our theoretical and numerical results in \S\ref{sSummary}.

\section{ Magnetic Stochasticity}\label{sMagnetic Stochasticity}

It is simple calculus to show that in the limit of vanishing magnetic diffusivity, the magnetic field becomes frozen into the fluid. Since diffusivity is indeed very small in most astrophysical systems, this mathematical result has led to the physical conclusion that in such situations magnetic field is frozen into the fluid as a good approximation. This phenomenon of "magnetic flux-freezing", also known as the Alfv\'en theorem, is usually applied in the laboratory and astrophysical fluids, implicitly assuming that MHD equations remain well-behaved. In the presence of turbulence, however, the velocity and magnetic fields would be generally singular and MHD equation ill-defined. Consequently, the Alfv\'en flux-freezing theorem does not generally apply in such environments.

In fact, for a magnetized fluid in the limit $\nu, \eta\rightarrow 0$, it turns out that even the very concept of magnetic field line encounters mathematical difficulties when the flow becomes turbulent. The existence and uniqueness of integral curves (field lines) is guaranteed only for Lipschitz continuous fields. Therefore, if the Lipschitz continuity condition is not satisfied, and hence uniqueness theorem cannot be applied, magnetic (and velocity) field lines would generally become ill-defined.

As discussed in \cite{JV2019}, difficulties similar to the diverging velocity and magnetic fields in turbulence were encountered in quantum electrodynamics (QED) and quantum chromodynamics (QCD) long time ago. For example, it turned out that calculating simple quantities such as mass or electric charge of particles leads to diverging expressions---the so-called Ultra-Violet (UV) divergences. This signals the fact that our theories, for instance QFT, are only approximations of nature valid only above a cut-off scale $l_{min}$ which is typically much larger than the Planck scale $l_P=(\hbar G/c^3)^{1/2}\sim 10^{-35}\;m$. As one resolution, lacking a complete theory describing nature down to very small scales of order the Planck length at present, Regularization and Renormalization Group (RG) methodologies were developed to resolve these theoretical difficulties. The general scheme of these formalisms are also applicable in many other fields including HD and MHD. Thus, one can remove the singularity of a given vector field, e.g., magnetic field, by coarse-graining or renormalizing the field. For a given velocity field ${\bf u(x}, t)$, or magnetic field ${\bf B(x}, t)$, for instance, the renormalized field at scale $l$ is simply the average field in a fluid parcel of size $l$ located at point $\bf x$ at time $t$, denoted by ${\bf u}_l({\bf x}, t)$ or .${\bf B}_l({\bf x}, t)$ Mathematically, this kind of averaging is called smoothing, which can be expressed in terms of generalized functions called distributions. In practice, to renormalize an arbitrary vector field ${\bf{B}}({\bf{x}}, t)$ at a length scale $l$, we can simply multiply it by a rapidly decaying function $G$ and integrate the result over the volume of interest:

\begin{equation}\label{1}
{\bf{B}}_l ({\bf{x}}, t)=\int_V G_l({\bf{r}})  {\bf{B}}({\bf{x+r}}, t) d^3r,
\end{equation}
where 

\begin{equation}\label{G}
G_l({\bf{r}}) =l^{-3} G\Big({{\bf r}\over l}\Big),
\end{equation}
with $G$ as a smooth, non-negative and rapidly decaying kernel. In fact, without loss of generality, we may assume

\begin{equation}\label{2}
G({\bf{r}})\geq 0,
\end{equation}

\begin{equation}\label{3}
\lim_{|\bf r|\rightarrow \infty} G({\bf{r}})\rightarrow 0,
\end{equation}

\begin{equation}\label{4}
\int_V d^3r G({\bf{r}})=1,
\end{equation}

\begin{equation}\label{5}
\int_V d^3r \; {\bf{r}}\;G({\bf{r}})=0,
\end{equation}
and
\begin{equation}
\int_V d^3r |{\bf{r}}|^2 \;G({\bf{r}})= 1.
\end{equation}
 We may also take $G({\bf{r}})=G(r)$ with $|{\bf{r}}|=r$, i.e. isotropic kernel, which leads to $\int d^3r\;r_i r_j G({\bf{r}})=\delta_{ij}/3$ \citep*{Eyink2018}. The renormalized field ${\bf{u}}_l$ represents the average field in a parcel of fluid of length scale $l$ at position $\bf x$.

\subsection{Stochasticity Level}\label{rev}

The scale-split energy density, $\psi({\bf{x}}, {\bf{r}}; t)$, is defined \citep*{JV2019} in terms of the renormalized vector field ${\bf{B}}_l({\bf{x}}, t)$ at scale $l$ and the renormalized field ${\bf{B}}_L({\bf{x}}, t)$ at scale $L$ as
\begin{equation}
\psi({\bf{x}}, {\bf{r}}, t)={1\over 2} \;{\bf{B}}_l({\bf{x}}, t){\bf{.B}}_L({\bf{x+r}}, t).
\end{equation}
Here we will be concerned only with $\psi({\bf{x}}, {\bf{r}}=0, t)\equiv  \psi({\bf{x}}, t)$. We write $\psi({\bf x}, t)=\phi({\bf x}, t)\chi({\bf x}, t)$ using the scalar fields
 
 \begin{equation}\label{phichi1}
\phi({\bf{x}}, t)=\begin{cases}
\hat{\bf{B}}_l({\bf{x}}, t).\hat{{\bf{B}}}_L({\bf{x}}, t) \;\;\;\;\;\;B_L\neq 0\;\&\;B_l\neq0,\\
0\;\;\;\;\;\;\;\;\;\;\;\;\;\;\;\;\;\;\;\;\;\;\;\;\;\;\;\;\;\;\;\;\;otherwise.
\end{cases}
\end{equation}
which is called the topology field 
and

\begin{equation}\label{phichi2}
\chi ({\bf{x}}, t)={1\over 2} B_l ({\bf{x}}, t) B_L({\bf{x}}, t).
\end{equation}
which is called cross energy field.

The stochasticity level $S_2$, topological deformation $T_2=\partial_t S_2(t)$, mean cross energy density $E_2(t)$, and field dissipation $D_2=\partial_t E_2(t)$ are given by (for more general definitions see \citep*{JV2019})

\begin{equation}\label{formulae}
S_2(t)={1\over 2} (1-\phi  )_{rms},
\end{equation}

\begin{equation}\label{Tdeform}
T_2(t)=  {1\over 4 S_2(t)} \int_V \;(\phi-1){\partial \phi\over \partial t}\; {d^3x\over V},
\end{equation}

\begin{equation}
E_2(t)=\chi_{rms},
\end{equation}

and
\begin{equation}
D_2(t)={1\over  E_2(t)}\int_V \chi \partial_t \chi{d^3x\over V}.
\end{equation}

It is easy to show \citep*{JV2019} that 

\begin{eqnarray}\notag
T_2(t)=&&{1\over 4 S_2(t)}  \int_V   \Big[\hat{\bf B}_l.\hat{\bf B}_L-1 \Big] \; \Big[  \hat{\bf{B}}_L. \Big({\partial_t {\bf B}_l \over B_l} \Big)_{\perp {\bf B}_l }  \\\label{T1}
&& + \hat{\bf{B}}_l. \Big({\partial_t {\bf B}_L\over B_L}  \Big)_{\perp{\bf B}_L }          \Big]  \;{d^3x\over V}.
\end{eqnarray}

Here, $(\;)_{\perp{\bf{B}}}$ represents the perpendicular component with respect to $\bf{B}$. In a similar way, we find

\begin{eqnarray}\notag
D_2(t)&=&{1\over E_2(t)}\int_V  \Big({B_l B_L\over 2}\Big)^2\\\label{Edissipation300}
&\times&  \Big[{\partial_t (B_L^2/2)\over B_L^2}+{\partial_t (B_l^2/2)\over B_l^2}    \Big]{d^3x\over V}.
\end{eqnarray}

For the magnetic field $\bf B$ in an electrically conducting fluid, the time derivative of the field appearing in these equations obeys the renormalized induction equation:
\begin{equation}
{\partial {\bf{B}}_l\over \partial t}=\nabla\times ( {\bf{u\times B}} )_l  - \nabla\times {\bf{P}}_l,
\end{equation}
where we have used the renormalized Ohm's law:
\begin{equation}
{\bf E}_l+({\bf u\times B})_l={\bf P}_l.
\end{equation}
Here $\bf P$ represents any non-ideal term in the generalized Ohm's law, e.g., the resistive electric field ${\bf P}=\eta \bf J$ with ${\bf J}=\nabla\times B$. This form of renormalized Ohm's law can also be re-written as
\begin{equation}\label{ROhm}
{\bf{E}}_l={\bf P}_l+{\bf R}_l-{\bf{ u}}_l \times {\bf{B}}_l.
\end{equation}
Therefore, even in the absence of any non-ideality $\bf P$, there is a non-linear term which is not necessarily negligible;
\begin{equation}\label{Rterm}
{\bf{R}}_l=-( {\bf{ u \times B}})_l+{\bf{u}}_l \times {\bf{B}}_l\equiv -{\cal{E}}_l.
\end{equation}
Here, the turbulent electric field (EMF) ${\cal{E}}_l\equiv-{\bf{R}}_l$ is the motional electric field induced by turbulent eddies of scales smaller than $l$ and plays a crucial role in magnetic dynamo theories. We find
\begin{equation}\label{induction1}
{\partial {\bf{B}}_l\over \partial t}=\nabla\times ( {\bf{u}}_l \times {\bf{B}}_l - {\bf{R}}_l-{\bf P}_l).
\end{equation}
We may assume that ${\bf P}_l$ is negligible in the inertial range of turbulence, which can basically be taken as the definition of the inertial range.

The only remaining piece to write down equations (\ref{T1}) and (\ref{Edissipation300}) is to note that the derivative of the unit vector $\hat{\bf{B}}={\bf{B}}/|{\bf{B}}|$ is associated with the perpendicular component of the induction equation:
  \begin{equation}\label{zap1}
  \partial_t \hat {\bf{B}}_l=\Big({\partial_t{\bf{B}}_l\over B_l} \Big)_\perp.
 \end{equation}
 
while the evolution of the magnitude of the magnetic field at scale $l$ is related to the parallel component of the induction equation:

\begin{equation}
{\partial B_l\over\partial t}=\Big({\partial {\bf B}_l\over\partial t}\Big)_\parallel.
\end{equation}

Putting all this together, we find

\begin{eqnarray}\notag
&&T_2(t)={1\over 4S_2(t)} \int_V \;{d^3x\over V}
 \;\underbrace{   \Big[ \; \hat{\bf B}_l.\hat{\bf B}_L-1 \Big] }_\text{self-entanglement (stochasticity)}\\\notag
 &\times&\Big[ \underbrace{  \Big({\hat{\bf{B}}_L\over B_l}. {\nabla\times ( {\bf{u}}_l \times {\bf{B}}_l )}_{\perp{\bf{B}}_l}+
{\hat{\bf{B}}_l\over B_L}. {\nabla\times ( {\bf{u}}_L \times {\bf{B}}_L ) }  _{\perp{\bf{B}}_L}  \Big)}_\text{turbulence (flow)}\\\label{stochasticity-rate2}
&&-\underbrace{  \Big( \hat{\bf{B}}_L.{\bf{\Sigma}}_l^\perp+\hat{\bf{B}}_l.{\bf{\Sigma}}^\perp_L  +\hat{\bf{B}}_L.{\boldsymbol{\sigma}}_l^\perp+\hat{\bf{B}}_l.{\boldsymbol{\sigma}}^\perp_L    \Big)}_\text{ slippage (reconnection)} \;\Big] ,
\end{eqnarray}

and

\begin{eqnarray}\notag
D_2(t) &=& {1\over E_2(t)}\int_V \Big({B_l B_L\over 2}\Big)^2 \\\notag
 &\times&\Big[ \underbrace{ \Big(  {\nabla\times({\bf u}_l\times{\bf B}_l)_{\parallel {\bf{B}}_l}\over B_l}+{\nabla\times({\bf u}_L\times{\bf B}_L)_{\parallel {\bf{B}}_L}\over B_L  }        \Big)}_\text{fluid-field interaction   }\\\label{Edissipation3001}
 &&- \underbrace{\Big( \Sigma_l^\parallel+\sigma_l^\parallel +\Sigma_L^\parallel+\sigma_L^\parallel    \Big)}_\text{magnetic dissipation}\Big]\;{d^3x\over V}.
\end{eqnarray}

In these equations, we have used the definitions

\begin{equation}\label{slipV}
{\bf{\Sigma}}_l={(\nabla\times{\bf{R}}_l) \over B_l},
\end{equation}
and
\begin{equation}\label{slipV2}
{\boldsymbol{\sigma}}_l={(\nabla\times{\bf{P}}_l) \over B_l},
\end{equation}
which are, respectively, the velocity-source terms in the turbulent inertial range and dissipative range. It has been already shown by Eyink \citep*{Eyink2015} that the perpendicular component of these vector fields (with respect to the magnetic field at the same scale), i.e., ${\bf{\Sigma}}^\perp_l$ or ${\boldsymbol{\sigma}}^\perp_l$ at small scales, are also the source terms driving the relative field-fluid velocity; see eq.(\ref{Eyink10}) in the next sub-section. Thus magnetic reconnection is intimately related (see \citep*{Eyink2015}; \citep*{JV2019}) to ${\bf{\Sigma}}^\perp_l\neq 0$ (${\boldsymbol{\sigma}}^\perp_l\neq 0$ at small scales). We will briefly review the slippage between magnetic field and the fluid in the next sub-section.

In passing, we also note that one can use the identity $\nabla\times({\bf u\times B})={\bf B.\nabla u-B\nabla.u-u.\nabla B+u\nabla.B}$ to write the bare induction equation as $D_t{\bf B}={\bf B.\nabla u-B\nabla. u}+\lambda \nabla^2 {\bf B}$ with Lagrangian derivative $D_t\equiv (\partial_t+{\bf u.\nabla}$). This is because in the "ideal MHD",  the magnetic diffusivity $\lambda$ tends to zero, $\lambda\rightarrow 0$, while the equations are assumed to be still well-defined. Using the continuity equation $D_t\rho+\rho\nabla.{\bf u}=0$, one finds

\begin{equation}\label{FF}
D_t\Big({ {\bf B}\over \rho}\Big)=\Big( {{\bf B}\over \rho}\Big).\nabla{\bf u}.
\end{equation}

This is the conventional flux freezing theorem presuming that MHD equations remain well-behaved in the limit $\lambda\rightarrow 0$ and the integral curves of ${\bf B}/\rho$ are advected with the fluid. For incompressible flow, the above expression become $D_t {\bf B}={\bf B}.\nabla{\bf u}$. Now if one tries to obtain these well-known results using the coarse-grained induction equation, eq.(\ref{induction1}), one finds for incompressible flow $D_t {\bf B}_l={\bf B}_l.\nabla{\bf u}_l-\nabla\times({\bf R}_l+{\bf P}_l)$. This expression indicates that flux-freezing does not hold in turbulence even in the limit of vanishing non-idealities ${\bf P}_l\rightarrow 0$ (e.g., for a vanishing resistive electric field ${\bf P}_l=\lambda \nabla\times{\bf B}_l\rightarrow 0$) and nonlinearities; ${\bf R}_l\rightarrow 0$. Instead, in addition to $\nabla\times{\bf P}_l \rightarrow 0$, magnetic flux freezing would more importantly also require $\nabla\times{\bf R}_l\rightarrow 0$ which generally does not hold in turbulence. These conditions, of course, can be expressed in terms of velocity-source terms defined by (\ref{slipV}) and  (\ref{slipV2}).

Note that the evolution of the direction vector of magnetic field, $\hat{\bf B}_l={ {\bf B}_l \over B_l }$, given by eq.(\ref{zap1}), is governed by  ${\bf{\Sigma}}^\perp_l$ and ${\boldsymbol{\sigma}}^\perp_l$ at small scales;

  \begin{equation}\label{zap10}
  \partial_t \hat {\bf{B}}_l={\nabla\times({\bf u}_l\times{\bf B}_l)_\perp \over B_l}-{\bf{\Sigma}}^\perp_l-{\boldsymbol{\sigma}}^\perp_l.
 \end{equation}

Let us summarize the implications of the above arguments about reconnection. Note that magnetic reconnection is a multi-scale phenomenon, and it occurs on a wide range of scales in a turbulent system. The renormalized Ohm's law has a collection of different terms with different physical meanings. The non-ideal effects in the Ohm's law, denoted collectively by ${\bf P}_l$ at scale $l$, arise from micro-scale plasma effects such as the resistive electric field, Hall effect etc. Such mechanisms drive reconnection at small diffusive scales. Such non-idealities, as discussed before, are mathematically represented by $\nabla\times{\bf P}_l$ in the induction equation. This term is also related to the velocity-source term for the magnetic field lines slipping through the fluid as we showed before. The width of reconnection zone is set by these small scales effects, e.g., by resistivity. On the other hand, the non-linear term ${\bf R}_l$ (at scale $l$) in the coarse-grained, generalized Ohm's law which arises from non-linear interactions below the arbitrary scale $l$. This is the same (with a negative sign) as the turbulent EMF in dynamo theories. At larger scales in the inertial range, $\Sigma_l$ dominates $\sigma_l$, which is negligible. At smaller scales, the resistive electric field drives the reconnection. However, at larger scales in the inertial range, the turbulent EMF dominates the resistive electric field in driving reconnection (\cite{Eyink2011}; \cite{Eyinketal.2013}; \cite{Eyink2015}). 

As we go down to smaller scales in the inertial range, $\Sigma_l$ decreases and eventually becomes comparable to $\sigma_l$ at the dissipative scale. Below the dissipative scale, $\sigma_l$ dominates $\Sigma_l$. Physically, all this translates into the well-known fact discussed in many papers in the last decade that reconnection occurs on all scales (\cite{Eyinketal.2013}; \cite{Eyink2015}; \cite{JV2019}). At smaller dissipative scales, it is driven by non-idealities denoted by ${\bf P}_l$, e.g., resistive electric field, whereas at larger scales in the inertial range it is driven by non-linearities denoted in the Ohm's law by ${\bf R}_l$, which are introduced by the turbulence. The explosive nature of super-linear Richardson diffusion brings distant field lines to small separations set by resistivity where they may reconnect while it also causes explosive separations between initially close field lines. These ideas are the essence of stochastic reconnection \cite{LazarianandVishniac1999}, general turbulent reconnection \cite{Eyink2015}, and stochastic flux freezing \cite{Eyink2011}. For example, \cite{Lalescuetal.2015} showed that the reconnection zone may in fact contain a great many current sheets instead of just one. This work shows that one can have a distribution of many current sheets. Also, the example studied in Figures (3-7) of \cite{Eyink2015} presents evidence for a very large-scale reconnection at the heliospheric current sheet (HCS).

\subsection{Field-Fluid Slippage }

Magnetic field in a turbulent, highly conducting fluid, e.g., a plasma, cannot be assumed perfectly frozen into the fluid as we discussed before. Instead the field may "slip" through the fluid. In order to quantify this field-fluid slippage mathematically, let us, following Eyink \citep*{Eyink2015}, denote by ${\boldsymbol{\xi}}(s; {\bf{x}}, t)$ an arbitrary point on the magnetic field line at time $t$ located at a distance $s$ from a base point $\bf x$ (along the field line), the unit tangent vector to the curve parametrized by $s$ is

\begin{equation}
{d\over ds} {\boldsymbol{\xi}}(s; {\bf{x}}, t)=\hat{\bf{B}}({\boldsymbol{\xi}}(s; {\bf{x}}, t), t), \;\;{\boldsymbol{\xi}}(s=0; {\bf{x}}, t)=\bf{x},
\end{equation}

where $\hat {\bf{B}}={\bf{B}}/|{\bf{B}}|$. On the other hand, the position of a fluid particle, which starts at ${\bf{x}}_0$ at time $t_0$ at a later time $t$ is governed by

\begin{equation}
{d\over dt} {\bf{x}}(t, {\bf{x}}_0, t_0)={\bf{u}}({\bf{x}}(t, {\bf{x}}_0, t_0), \;\; {\bf{x}}(t_0, {\bf{x}}_0, t_0)={\bf{x}}_0.
\end{equation}

If magnetic flux-freezing holds, we should be able to parametrize both field lines and the trajectories of the fluid particles together using the same function ${\boldsymbol{\xi}}\equiv {\bf{x}}$. In other words, in that case, we could find a function $s(t, s_0, x_0)$ such that  ${\boldsymbol{\xi}}(s(t; s_0, {\bf{x}}_0); {\bf{x}}(t; {\bf{x}}_0, t_0), t)={\bf{x}}(t; {\boldsymbol{\xi}}(s_0; {\bf{x}}_0, t_0), t_0)$. The derivative of this equation reveals that the flux freezing condition, $(d/dt) {\boldsymbol{\xi}}={\bf{u}}({\boldsymbol{\xi}}, t)\equiv \tilde {\bf{u}}$, holds if and only if

\begin{equation}
\dot s (t) \hat {\bf{B}} ({\boldsymbol{\xi}}, t)+D_t {\boldsymbol{\xi}}=\tilde {\bf{u}},
\end{equation}

where $D_t=\partial_t+{\bf{u}}.\nabla$ is the convective derivative. To determine $s(t)$, we can write
\begin{equation}
\dot s (t)=(\tilde{\bf{u}}-D_t {\boldsymbol{\xi}}).\hat{\bf{B}}=(\tilde{\bf{u}}-D_t{\boldsymbol{\xi}})_\parallel,\;\;s(t_0)=s_0.
\end{equation}

Consequently, the condition $d{\boldsymbol{\xi}}/dt=\tilde{\bf{u}}$ will hold if and only if for all $s$, $\bf x$ and $t$,

\begin{equation}
(D_t{\boldsymbol{\xi}})_\perp (s; {\bf{x}}, t)-{\bf{u}}_\perp ({\boldsymbol{\xi}}(s; {\bf{x}}, t), t)=0.
\end{equation}

This expression is another way to quantify flux-freezing. It states that the relative perpendicular velocity  (with respect to the field line) between the field line and fluid elements vanishes. Thus when flux freezing condition is not satisfied, this relative velocity has a non-zero value which we denote by

\begin{equation}
\Delta {\bf{w}}_\perp (s; {\bf{x}}, t)=(D_t{\boldsymbol{\xi}}-\tilde {\bf{u}})_\perp (s; {\bf{x}}, t).
\end{equation}
Therefore flux-freezing condition translates into $\Delta{\bf{w}}_\perp\equiv 0$. It is easy to show (for details see \citep*{Eyink2015}) that

\begin{equation}\label{Eyink10}
{d\over ds}\Delta {\bf{w}}_\perp-\Big[ ( \nabla_{\boldsymbol{\xi}}\hat{\bf{B}})^T-(\hat{\bf{B}}\hat{\bf{B}}).(\nabla_{\boldsymbol{\xi}}\hat{\bf{B}})          \Big].\Delta{\bf{w}}_\perp=-{(\nabla\times{\bf{P}})_\perp\over |{\bf{B}}|}.
\end{equation}

Hence, assuming that the field remains smooth as ${\bf{P}}\rightarrow 0$, one might naively conclude that flux freezing holds and the field lines move with the fluid elements with no slippage. In fact, the above expression indicates that flux freezing holds if $\hat{\bf{B}}\times (\nabla\times {\bf{P}})=0$. This condition has long been known as the general condition for flux freezing \citep*{Newcomb1958}: $(\nabla\times {\bf P})_\parallel=0$. Note that this conclusion, in the limit ${\bf P}\rightarrow 0$, is based on the assumption that magnetic field remains smooth and differentiable. We also emphasize that the source term in eq.(\ref{Eyink10}) is the same slip-velocity source term given by eq.(\ref{slipV2}) which is related to the field topology through eq.(\ref{zap10}); for a detailed mathematical treatment of this relationship see \cite{JV2019}.

\subsection{Magnetic Reconnection}\label{sreconnection}

In a typical reconnection event, two regions sharing a boundary with intense magnetic shear (usually called a current sheet as a large magnetic shear indicates large electric currents) are pushed toward each other with a reconnection speed $V_R$. Because of mass conservation, matter is then ejected with a fraction of the local Alfv\'en speed $V_A$. In order to estimate the latter, one can assume that the magnetic energy $B^2/2$ is totally converted to the kinetic energy of the outflow which moves with velocity $u_x$;
\begin{equation}
\rho u_x^2\simeq B^2/2,
\end{equation}
where $\rho$ is the density. This leads to an ejection speed of order the local Alfv\'en speed, $u_x \simeq V_A$. As for the inflow or reconnection speed, one can start with the Ohm's law 
\begin{equation}\label{Ohm1}
 {\bf{E}}+{\bf{u\times B}}=\eta {\bf{J}},
 \end{equation}
 where $\eta$ is the diffusivity, ${\bf E}$ the electric field, $\bf u$ the velocity field and $\bf J=\nabla\times B$ the electric current.
As an order of magnitude scaling, the above result leads to $J \sim  {V_RB}/{\eta}$. Note that the term $\eta \bf J$ in the Ohm's law becomes important because a large current $\bf{J=\nabla\times B}$ forms as a result of large magnetic field gradient (shear) while the diffusivity is typically very small. Thus, energy loss due to Ohmic dissipation, $\eta \int J^2 d^3x$, is appreciable only if there are very large magnetic field gradients in the volume. In general, reconnection requires only a small, but finite, diffusivity to proceed.

For a current sheet of thickness $ \delta$ and length $\Delta$, in the steady state, one can apply the Amp\'ere's law to estimate the current, $J \sim {B}/{\delta}$ and thus we get $V_R\sim {\eta}/{\delta}$. In order to use energy conservation in a reconnection zone of length $\Delta$ and width $\delta$, we note that the Poynting energy flux into the zone is $V_R B^2\Delta$. This energy is consumed in two ways: Ohmic dissipation $J^2 \eta \delta \Delta $ and the kinetic energy flux of the outflow $\rho V_A^2 (V_A \delta)$ \citep*{Zweibeletal.2016}. We find
\begin{equation}
V_A^3 \delta=V_{R} V_A^2\Delta-\epsilon \Delta\delta,
\end{equation}
where $\epsilon={\bf{E.J}}/\rho$ is the energy dissipation rate. Neglecting the dissipation, the last term, we would recover the mass conservation $V_A \delta=V_{R} \Delta$. Putting all this together, we obtain a reconnection speed of order

\begin{equation} \label{15}
V_R=\left  ( {\eta \frac{V_A}{\Delta}}\right)^{1/2}=V_AS^{-1/2},
\end{equation}

 where $S=V_A\Delta/ \eta$ is the Lundquist number. Note that the Sweet-Parker (\citep*{Parker1957}; \citep*{Sweet1958}) time scale $t_R=\delta\Delta/\eta$ is shorter than the resistive time scale $t_\eta=\Delta^2/\eta$ by a factor of $\sqrt{S}$; $t_R=t_\eta/\sqrt{S}$ and longer than the Alfv\'en time scale $t_A=\Delta/V_A$ by the same factor; $t_R=\sqrt{S} t_A$. In the solar corona, where $S$ is of order $10^{12}$, the above expression leads to a reconnection time of order $t_R \geq 10^6\;s$. However, the measured time scale is of order $t_R \sim 100\;s$. For instance, the field topology in the soft-x-ray pictures changes in a time scale of minutes or at most hours which is much shorter than the Sweet- Parker time. Thus, in spite of the fact that the Sweet-Parker scheme predicts much faster conversion rate for magnetic energy than the global diffusion, nevertheless, it is still much too slow compared with the observations \cite{Yamadaetal.2010}. Also note that with vanishing diffusivity, the width of the current sheet vanishes as well, and reconnection may only proceed with an anomalous diffusivity discussed below (see also \cite{Shayetal.2001}).
 
 It turns out that although Sweet-Parker model is a good approximation in laminar flows where magnetic flux tubes undergo large scale Taylor (normal) diffusion, however, it fails utterly in turbulent systems as expected because it ignores all turbulent effects on magnetic field and the flow. In fact, magnetic flux freezing breaks down in turbulence and Lagrangian particle trajectories become random. This leads to stochasticity in magnetic fields in turbulence for which a generalized version of flux freezing, stochastic flux freezing, applies instead of conventional Alfv\'en theorem. In the next section, we quantify magnetic stochasticity and briefly explain its relationship with magnetic topology.

 \subsection{Stochastic Reconnection}

 The Sweet-Parker scheme can also be understood in terms of magnetic field diffusion. On very large scales, magnetic flux tubes diffuse away as a result of magnetic diffusivity. In hydrodynamic turbulence,Taylor diffusion (the linear diffusion present also in Brownian motion) indicates that the average (rms) distance of a particle from a fixed point, $y(t)$, increases with time $t$ as 
\begin{equation}\label{NormalDiff}
y^2(t)\simeq D t,
\end{equation}
where $D$ is diffusion coefficient. This is normal (Taylor) diffusion in which average square distance between a particle (a dye molecule in water) and a fixed point increases linearly with time; $y^2\propto t$. The normal diffusion of magnetic field follows a mathematically similar relationship between the rms distance (spreading) between magnetic field lines instead of particle separation (see e.g., \cite{Eyinketal.2013}; \cite{JafariandVishniac2018}). In this case, the diffusion coefficient is the magnetic diffusivity $\eta$. Whether the medium is turbulent or not, this diffusion scheme, for fluid particles in hydrodynamic turbulence and magnetic field lines in magnetohydrodynamic turbulence, will apply but with different diffusion coefficients. In other words, turbulence will increase the diffusion coefficient making the diffusion process faster and more efficient but the nature of diffusion is linear (in time) at scales much larger than the turbulent inertial range. 

The normal diffusion scheme cannot be used in the inertial range of turbulence (see below). In the absence of turbulence, in a reconnection zone with width $\delta$ and length $\Delta$ (parallel to the anti-parallel magnetic fields), substituting the Alfv\'en time scale $t_A=\Delta/V_A$ in eq.(\ref{NormalDiff}), and using mass conservation $ V_A y= V_R\Delta$, we recover the Sweet-Parker reconnection speed, given by eq.(\ref{15}). Therefore, Sweet-Parker reconnection can be valid only in the absence of turbulence.

Reconnection itself, along with other instabilities such as tearing modes \citep*{Furthetal.1963}, will generate turbulence (\citep*{Eastwoodetal.2009}; also see e.g., \citep*{JafariandVishniac2018} for a review of turbulent and non-turbulent reconnection models). In the turbulence inertial range, i.e., at scales larger than dissipative scale but much smaller than the larger scales where Taylor (normal) diffusion occurs, particles undergo super-linear Richardson diffusion; $\delta^2\propto t^3$. It is important to notice that Richardson diffusion is 2-particle diffusion, i.e., $\delta$ is the separation between two particles undergoing diffusion in the inertial range unlike $y(t)$ in eq.(\ref{NormalDiff}) which corresponds to (one-particle) Taylor diffusion. If we consider magnetic diffusion in the turbulence inertial range, we have to consider Richardson diffusion of the field lines. On these scales, therefore, the Sweet-Parker model obviously cannot be applied. The Richardson probability density for particle separation vector ${\bf{l}}={\bf{x}}_1-{\bf{x}}_2$, with a scale-dependent diffusion  coefficient $K(l)\sim K_0 l^{4/3}$, satisfies $\partial_t P({\bf{l}},t)=\nabla_{l_i}\Big( K(l) \nabla_{l_i} P({\bf{l}},t)\Big)$ with a similarity solution \citep*{Eyink2011},

\begin{equation}\label{Richardson1}
P({\bf{l}},t)={A\over (K_0 t)^{9/2} }\exp{\Big( -{9l^{2/3}\over 4 K_0 t} \Big) }.
\end{equation}
Using this probability density to average $l^2$, one finds $\langle l^2(t)\rangle=(1144/81) K_0^3 \; t^3$. This is intimately related to Kolmogorov's relation 
\begin{equation}\label{t3}
l^2(t)\sim (g_0 \epsilon) t^3,
\end{equation} 
which is a solution to the initial value problem $dl(t)/dt=\delta u(l)=(3/2)(g_0\epsilon l)^{1/3}$, $l(0)=l_0$ for sufficiently long times $t\gg t_0$. Here $g_0$ is Richardson-Obukhov constant and $\epsilon$ the mean energy dissipation rate.

The results implied by eq.(\ref{t3}) can also be obtained using a simple dimensional analysis. In the inertial range of the turbulent cascade, \citep*{Kolmogorov1941}, the eddy turnover time is of order $t \sim \epsilon^{-1/3}\delta^{2/3}$ with $\delta$ being the length scale perpendicular to the mean magnetic field. Here, $\epsilon\simeq V_T^2 V_A/l_\parallel$ denotes the energy transfer rate, with turbulent velocity $V_T$ and parallel energy injection length scale $l_\parallel$. This corresponds to the Richardson diffusion:
\begin{equation}
\delta^2(t)\sim \epsilon t^3,\;\;\;\;\;\;\;(turbulent\;\;medium).
\end{equation}

A comparison of this expression with eq.(\ref{NormalDiff}) shows that the Richardson diffusion broadens the reconnection zone by faster spreading the field lines as it is a super-linear diffusion, $\delta^2\propto t^{3}$. Using mass conservation $ V_A\delta= V_R\Delta$, and substituting the Alfv\'en time $t_A=\Delta/V_A$, one arrives at the fast reconnection rate predicted in stochastic model (\citep*{LazarianandVishniac1999}; \citep*{Eyinketal.2011}; \citep*{Jafarietal.2018}; for a more detailed review see \citep*{JafariandVishniac2018}):
\begin{equation}\label{LV1999}
V_R\sim V_T \; Min \Big[ \Big( {\Delta\over l_\parallel}\Big)^{1/2}, \Big( {l_\parallel\over \Delta}\Big)^{1/2}\Big].
\end{equation}

Here, depending on the sizes of the current sheet $\Delta$ and the parallel energy injection scale $l_\parallel$, the smaller ratio, either $(\Delta/l_\parallel)^{1/2}$ or $(l_\parallel/\Delta)^{1/2}$, is taken. This reconnection speed is of order the large turbulent eddy velocity, is independent of diffusivity and is in agreement with numerical simulations to date (\citep*{Kowaletal.2009}; \citep*{Kowaletal.2012}). The stochastic model of reconnection was also examined with a large viscosity to diffusivity ratio in a recent work \citep*{Jafarietal.2018}.

\section{Stochasticity and Topology Change}\label{sStatistical Study of Stochasticity}

Turbulence in general will tend to tangle an initially smooth magnetic field, locally changing the magnetic field direction $\hat{\bf B}_l$ in a stochastic way. This effect corresponds to the term $\nabla\times({\bf u}_l\times{\bf B}_l)_\perp/B_l$ implicit in the parentheses on the RHS of eq.(\ref{zap1}), which reads
  \begin{equation}\notag
  \partial_t \hat {\bf{B}}_l=\Big({\partial_t{\bf{B}}_l\over B_l} \Big)_\perp.
 \end{equation}
 In terms of the field topology, this corresponds to the turbulence (flow) terms in  eq.(\ref{stochasticity-rate2}), which is
 \begin{eqnarray}\notag
&&T_2(t)={1\over 4S_2} \int_V \;{d^3x\over V}
 \;\underbrace{   \Big[ \; \hat{\bf B}_l.\hat{\bf B}_L-1 \Big] }_\text{self-entanglement (stochasticity)}\\\notag
 &\times&\Big[ \underbrace{  \Big({\hat{\bf{B}}_L\over B_l}. {\nabla\times ( {\bf{u}}_l \times {\bf{B}}_l )}_{\perp{\bf{B}}_l}+
{\hat{\bf{B}}_l\over B_L}. {\nabla\times ( {\bf{u}}_L \times {\bf{B}}_L ) }  _{\perp{\bf{B}}_L}  \Big)}_\text{turbulence (flow)}\\\notag
&&-\underbrace{  \Big( \hat{\bf{B}}_L.{\bf{\Sigma}}_l^\perp+\hat{\bf{B}}_l.{\bf{\Sigma}}^\perp_L  +\hat{\bf{B}}_L.{\boldsymbol{\sigma}}_l^\perp+\hat{\bf{B}}_l.{\boldsymbol{\sigma}}^\perp_L    \Big)}_\text{ slippage (reconnection)} \;\Big] ,
\end{eqnarray}
 This effect also makes $\hat{\bf B}_l$ deviate from $\hat{\bf B}_L$ thus the turbulence (flow) term in eq.(\ref{stochasticity-rate2}) will increase in magnitude. As a result, stochasticity level starts to increase, i.e., $T_2=\partial_t S_2\geq 0$, until the tangled field starts to resist more tangling and bending by slipping through the fluid. This effect is already known to be related to $\Sigma^\perp\neq 0$ (and $\sigma^\perp\neq 0$) whose effect is represented by the slippage (reconnection) term. This can lead to a sudden motion of the field lines relative to the fluid quickly decreasing the stochasticity level $T_2=\partial_t S_2\leq  0$. Therefore, at some point between these two stages, $T_2=\partial_t S_2=0$.

How is the field magnitude affected during this process? We note that the coarse-grained field ${\bf B}_l$ is in fact the average field in a spatial volume of scale $l$. To see this simple fact more clearly, we first note that point-wise we have

\begin{equation}\notag
|\phi({\bf x}, t)|\leq \Big|\int_V  {d^3r\over l^3} \int_V {d^3R\over L^3} \; \hat{\bf B}({\bf x+r}, t).\hat{\bf B}({\bf x+R}, t)  \Big|,
\end{equation}

which is, by definition, equal to unity for a smooth (non-tangled) field. To increase an initially vanishing stochasticity level ${1\over 2}(1-\phi)_{rms}=0$ to a non-zero value, the stochastic variable $\phi$ is to deviate from unity, i.e., the unit vector $\hat{\bf B}$ must in general take different directions at different points. On the other hand, it is simple calculus to see that
\begin{eqnarray}\notag
|{\bf B}_l({\bf x}, t)|&=& \Big| \int_V {d^3r\over l^3} G(r/l) {\bf B}({\bf x+r}, t)\Big|\\\notag
&\leq &\Big|\int_V G(r/l) {d^3r\over l^3} \Big|\;\Big| \int_V{\bf B}({\bf x+r}, t) {d^3r\over l^3}\Big|\\\notag
&\lesssim& \Big|  \sum_i {\bf B}({\bf x}+{\bf r}_i, t) \Big| \Big|  \sum_i {\Delta^3 r_i\over l^3} \Big|.
\end{eqnarray}

 \begin{figure}[h]
 \begin{centering}
\includegraphics[scale=.42]{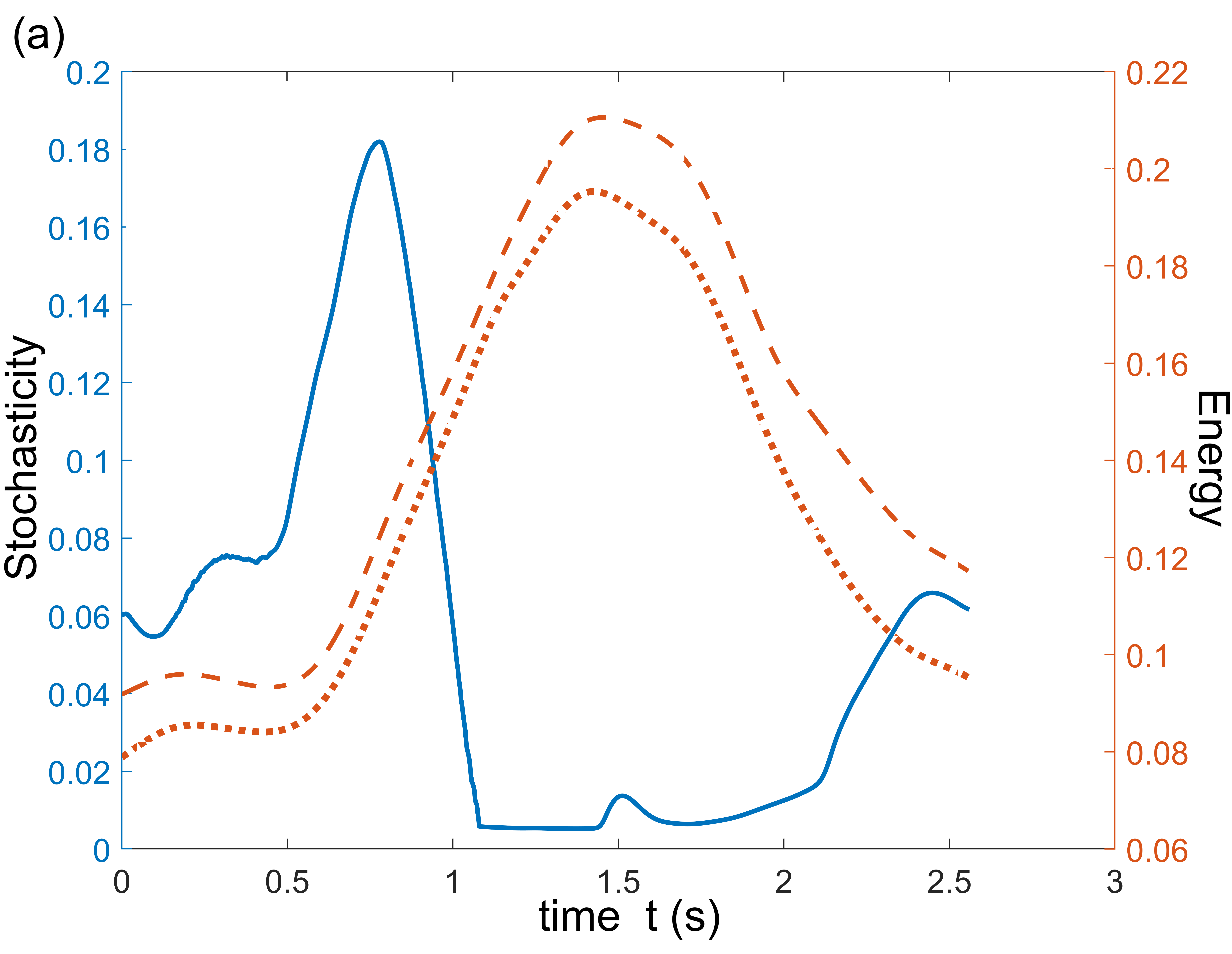}
\includegraphics[scale=.213]{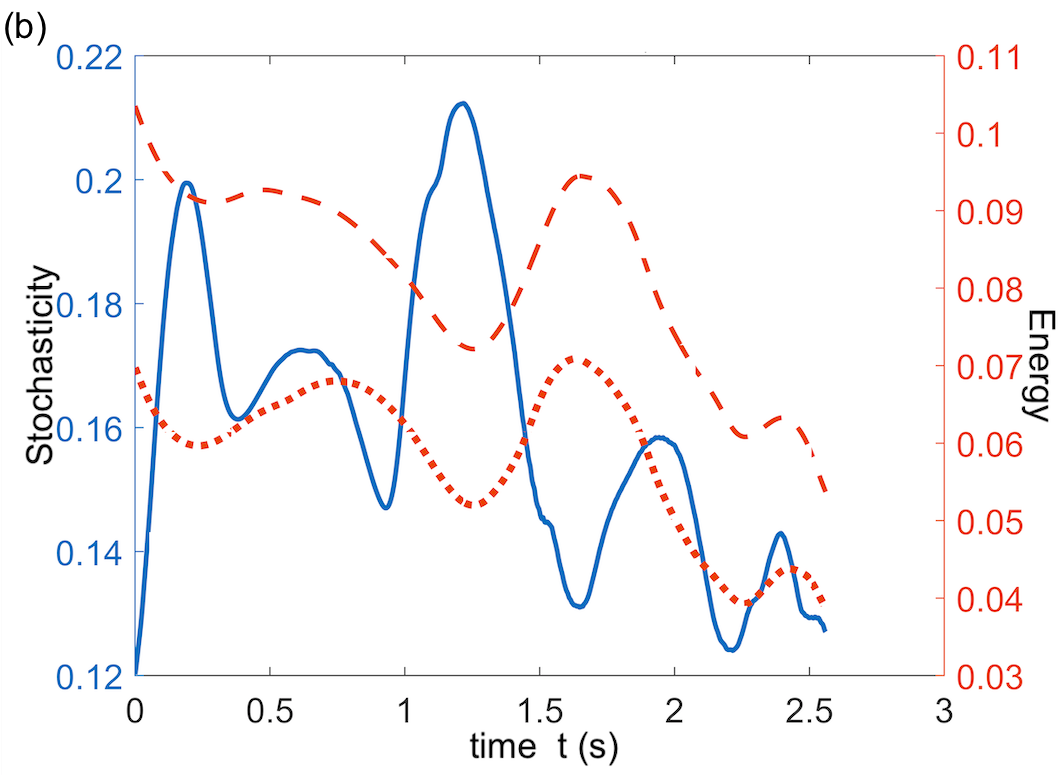}
\includegraphics[scale=.42]{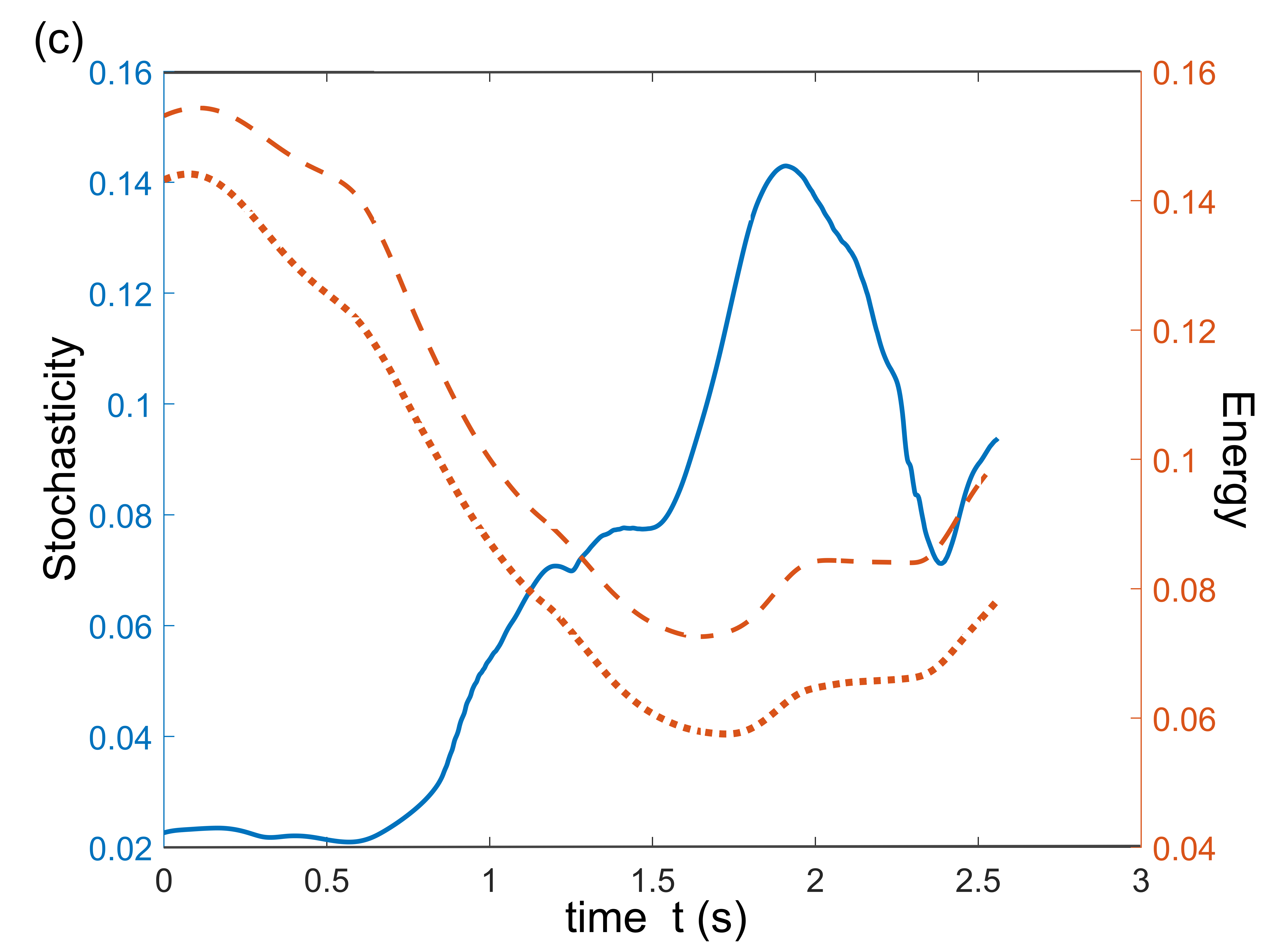}
\caption {\footnotesize {Magnetic stochasticity $S_2(t)$ (blue, solid curve), cross energy $E_2(t)$ (red, dotted curve) and mean magnetic energy density $(B^2/2)_{rms}$ (red, dashed curve) for three different sub-volumes of the simulation box. Apart from small, short-lived fluctuations, the overall trend is as theoretically expected. }}\label{zap1}
\end{centering}
\end{figure}
Hence, increasing stochasticity at scale $l$ makes local ${\bf B}$ fields at scales $d\leq l$ less aligned with one another which in turn partially cancel one another out when we average, i.e., coarse-grain, them to get the field ${\bf B}_L$ at a larger scale $L\geq l$. Thus the average $\int_VB_L^2/2$ will generally decrease by increasing stochasticity at smaller scales, that is $T_2=\partial_t S_2\geq  0$ leads to $\int_V \partial_t (B_l^2/2)\leq 0$ and $\int_V \partial_t (B_L^2/2)\leq 0$. (Also we note that during a field-fluid slippage at scale $l$, the kinetic energy of accelerating particles is extracted from $B_l^2/2$ which means $\int_V \partial_t (B_l^2/2)\leq 0$.) From eq.(\ref{Edissipation300}), it is easy to see that $\int_V \partial_t (B_l^2/2)\leq 0$ and $\int_V \partial_t (B_L^2/2)\leq 0$ indicate $D_2=\partial_t E_2\leq 0$. Similarly, a decreasing stochasticity, i.e., $T_2=\partial_t S_2\leq 0$, is accompanied with $\int_V\partial_t B_l^2\geq 0$ and $\int_V\partial_t(B_L^2/2)\geq 0$, hence $D_2=\partial_t E_2\geq 0$. At the peak of field-fluid slippage, therefore, $S_2$ reaches a maximum approximately followed by a minimum of $E_2(t)$.

In order to test the theoretical predictions discussed in the previous paragraph, and in the next sections, we use a homogeneous, incompressible MHD numerical simulation archived in an online, web-accessible database (\citep*{JHTDB};\citep*{JHTB1};\citep*{JHTB2}). This is a direct numerical simulation (DNS) from Johns Hopkins Turbulence Database, using $1024^3$ nodes, which solves incompressible MHD equations using pseudo-spectral method. The simulation time is $2.56\;s$ and $1024$ time-steps are available (the frames are stored at every $10$ time-steps of the DNS). We test our predictions in several randomly selected sub-volumes of the simulation box with different sizes. This reduces the computation time while it assures that the predicted features are prevalent everywhere in the box. 

Fig.(\ref{zap1}) represents magnetic stochasticity, cross energy and rms magnetic energy density in three randomly selected sub-volumes each of size $194\times 42\times 33$ in grid units, equivalent to $1.2\times 0.26\times 0.20$ in physical units. Similar results are obtained by repeating this computation in other, smaller and larger, randomly selected sub-volumes. The observed trends are typical and do not in general depend on the coordinates or sizes of the sub-volumes. We also note that although the predicted statistical relationship between magnetic stochasticity $S_2(t)$ and cross energy $E_2(t)$ is observed in all sub-volumes, however, the rate of change and the total change in magnetic energy can be very different in different sub-volumes. For instance, Fig.(\ref{quietSV}) plots magnetic stochasticity and cross energy in two sub-volumes in which the total change in cross energy $E_2(t)$ is very small compared to what is observed in Fig.(\ref{zap1}). This might indicate that in the sub-volumes corresponding to Fig.(\ref{quietSV}), only a simple slippage between the field and fluid has occurred without much magnetic energy dissipation whereas in Fig.(\ref{zap1}) both slippage (change in stochasticity) and energy dissipation (change in energy) are involved. This might suggest that reconnection involves both slippage and energy dissipation. This picture requires more investigations, however, we present such a hypothetical classification in Table.(\ref{table1}) in \S \ref{IIIC}.

\subsection{Field-Flow Interaction: A Toy Model}

\begin{figure}[h]
 \begin{centering}
\includegraphics[scale=.21]{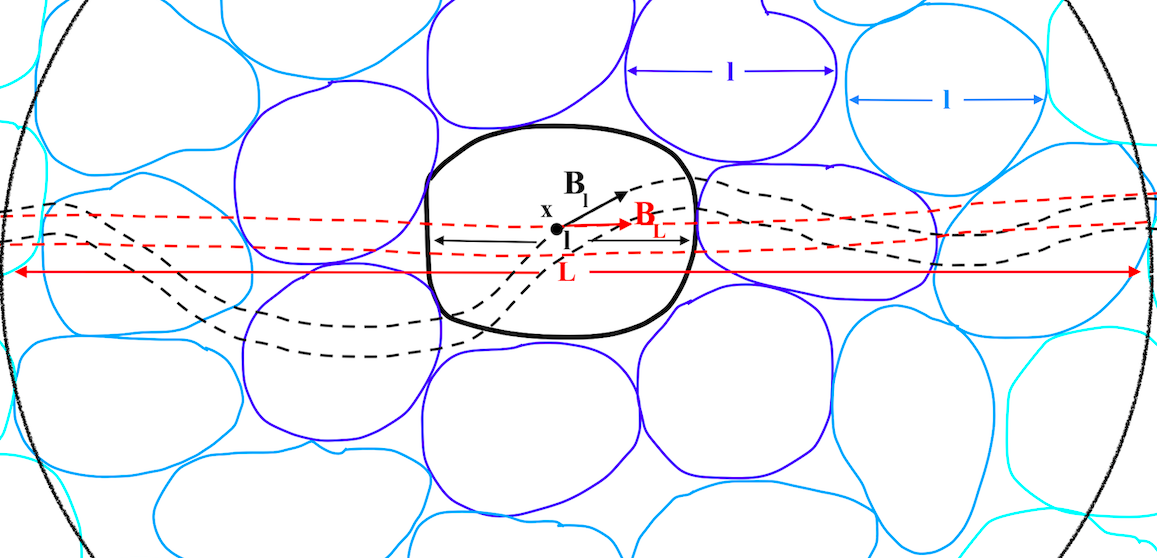}
\caption {\footnotesize {Fluid parcels of scale $l$ as a collection of classical spins of magnetic moment ${\boldsymbol{\mu}}_l\propto {\bf B}_l$, in the mean field approximation (classical Ising model).}}\label{Paramagnetism}
\end{centering}
\end{figure}

We can also make a simple toy model illustrating the points made above regarding the relationship between magnetic stochasticity and the kinetic energy of turbulence. Suppose we divide a given volume of fluid, of size $L$, into small sub-volumes of size $l\ll L$; see Fig.(\ref{Paramagnetism}). If one ignores all the fluid surrounding a parcel of fluid of scale $l$ at point $\bf x$, focusing only on the parcel itself, the coarse-grained magnetic field at point $\bf x$ inside the parcel would still be approximately given by ${\bf B}_l$. That is to say, the contribution of outer points, at distances $\gg l$ is negligible in getting the coarse-grained field ${\bf B}_l$ inside the parcel. On the other hand, had we instead eliminated the fluid parcel of scale $\sim l$ around point $\bf x$, retaining the rest of the fluid in the surrounding, the coarse-grained field at $\bf x$ over a large scale $L\gg l$, would be still ${\bf B}_L$ within a good accuracy. 

This is the motivation to divide the total volume $V\sim L^3$ into regions of scale $l\ll L$ and consider each region as a classical spin with magnetic field ${\bf B}_l$ and magnetic moment ${\boldsymbol{\mu}}_l\propto l^3 {\bf B}_l$ embedded in the mean field ${\bf B}_L$ generated by the neighboring fluid parcels. Hence a typical parcel will possess a magnetic energy $-{\boldsymbol\mu}_l.{\bf B}_L\propto -l^3 {\bf B}_l.{\bf B}_L$. We can also define a "temperature" $T_l({\bf x}, t)$ proportional to the average available kinetic energy at scale $l$, which we denote by $v_{l, L}^2$. Thus the Boltzmann's $\beta$-factor may be defined as\footnote{Thermodynamic equilibrium in this context translates into homogeneity and isotropy which is unrealistic in MHD turbulence. We work in the weak field regime $B_l B_L\ll v^2_{l, L}$ and use this approximation only as an instructive toy model.} 

\begin{equation}\label{beta}
\beta_l:={1\over v^2_{l, L}}.
\end{equation}
In terms of the scale-split magnetic energy density, $\psi({\bf x}, t, \theta({\bf x}, t))={1\over 2} {\bf B}_l.{\bf B}_L={1\over 2}B_lB_L\cos\theta$, the Boltzmann factor becomes $e^{\beta ({\bf x}, t) \psi({\bf x}, t, \theta_i({\bf x}, t))}$, and therefore the partition function is ${\cal{Z}}=\sum_i e^{\beta_l ({\bf x}, t) \psi({\bf x}, t, \theta_i({\bf x}, t))}$. More generally, we can attribute a magnetic moment ${\boldsymbol{\mu}}_l=g {\bf B}_l$, with a constant $g\propto l^3 $, to a fluid parcel of scale $l$ which leads to the partition function 

\begin{equation}
{\cal{Z}}=\int d\cos\theta \;e^{g \beta_l  \psi }.
\end{equation}

\begin{figure}[t]
 \begin{centering}
\includegraphics[scale=.32]{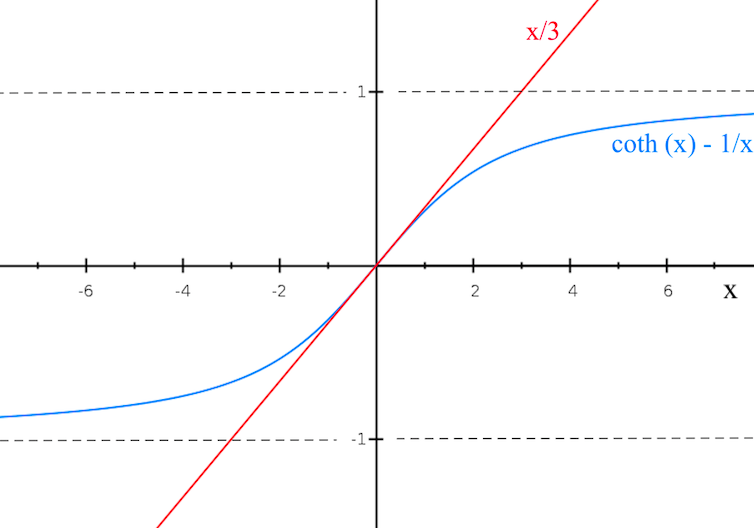}
\caption {\footnotesize {The ensemble average $\overline \phi=\overline{(\hat{\bf B}_l.\hat{\bf B}_L)}$ is given by eq.(\ref{ensemblephi}). This function also appears in the similar problem of finding the average magnetic moment in a collection of classical spins in statistical mechanics (also closely related to a classical version of the Ising model in weak field approximation). With $x=B_lB_L/v_{l,L}^2 $, defined in eq.(\ref{beta}), this is the function $g(x)=\coth x-1/x$. In the weak field regime, $x\ll 1$, this function is approximated linearly by $x/3$.}}\label{coth2}
\end{centering}
\end{figure}

 If we absorb the proportionality constant $g$ into the definition of turbulent kinetic energy, i.e., the $\beta$ factor, the probability of finding a region of scale $l$ whose magnetic field ${\bf B}_l$ makes an angle between $\theta$ and $\theta+d\theta$ with the large scale field ${\bf B}_L$ is given by

\begin{equation}\notag
p({\bf x}, t; \theta({\bf x}, t))d\theta={e^{{\beta_l B_lB_L}\cos\theta}d\cos\theta \over \int_{-1}^{+1} d\cos\theta e^{{\beta_l B_lB_L}\cos\theta}}.
\end{equation}
The "ensemble" average\footnote{Throughout this paper, we avoid using ensemble averages and instead we rely only on one single realization of the velocity and magnetic fields. However, we make an exception here since this simple toy model is best related to paramagnetism using ensemble averaging.}of $\phi=\hat{\bf B}_l.\hat{\bf B}_L=\cos\theta$ is given by

\begin{equation}\label{ensemblephi}
\overline\phi\simeq \coth\Big( {B_lB_L\over v_{l,L}^2} \Big)-\Big( {v_{l,L}^2 \over B_lB_L}\Big).
\end{equation}
This expression in the weak field limit i.e., $B_lB_L\ll v_{l,L}^2$, becomes
\begin{equation}\notag
\overline\phi\simeq {1\over 3}{B_l B_L\over v_{l,L}^2 }={1\over 3}{B_l^2\over v_{l,L}^2 }{B_L\over B_l},
\end{equation}
where we have used the approximation $\coth x\simeq x/3$ for small $x$; see Fig.(\ref{coth2}).

In terms of $\chi={1\over 2} B_lB_L$, therefore, $\overline\phi\simeq {2\over 3} {\chi\over \langle u_l^2\rangle}$. We have also

\begin{equation}
{1-\overline\phi\over 2} ={1\over 2}\Big(1- {1\over 3}{B_l^2\over v_{l,L}^2 }{B_L\over B_l} \Big).
\end{equation}

\begin{figure}[t]
 \begin{centering}
\includegraphics[scale=.225]{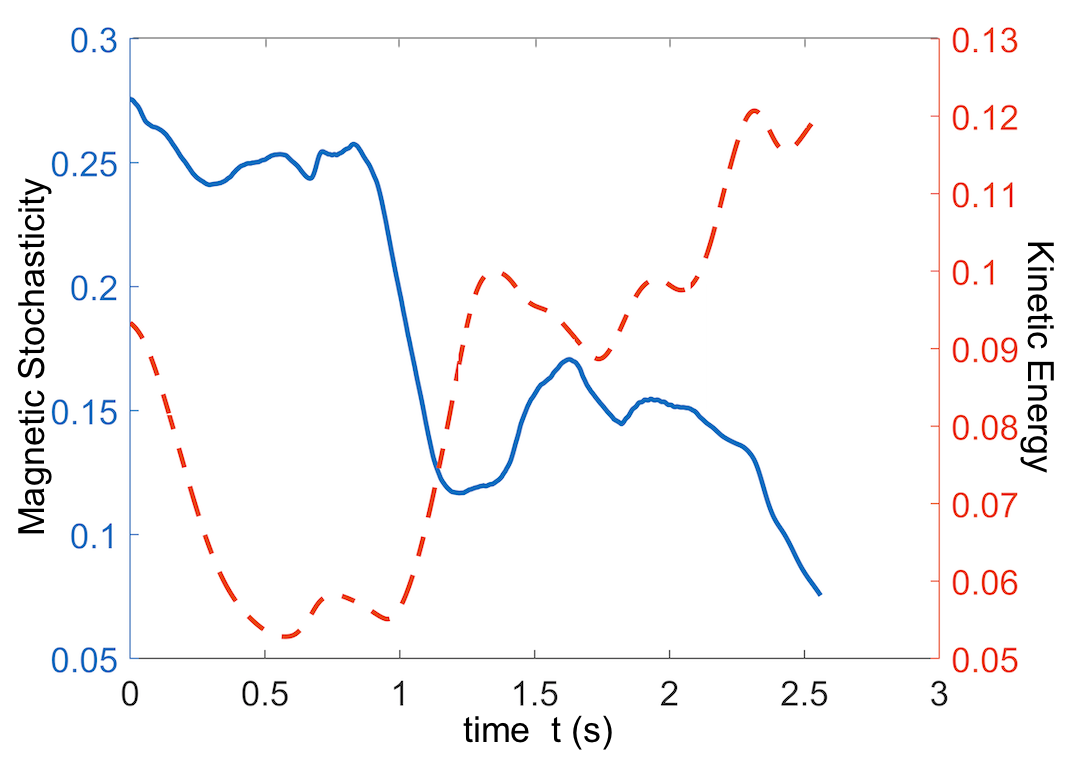}
\caption {\footnotesize {Magnetic stochasticity level $S_2(t)$ (blue, solid curve) and turbulent kinetic energy $(u^2/2)_{rms}$ (red, dashed curve), which is taken as a substitute for $v_{l, L}^2$ for simplicity. In the numerical simulation used to obtain this graph and similar other ones in different sub-volumes, the weak-field condition $B_lB_L\ll v^2_{l, L}\leq u_{rms}^2$ is not satisfied, and thus we are not certainly in the weak field regime to use eq.(\ref{stochasticity}). Still, in several sub-volumes of the simulation box, the theoretical expectation predicted by eq.(\ref{stochasticity}) is observed: magnetic stochasticity increases (decreases) as the turbulent kinetic energy increases (decreases). Despite this partial agreement, however, the relationship between stochasticity and turbulent kinetic energy remains speculative and in need of more numerical studies.  }}\label{sub1}
\end{centering}
\end{figure}

For small variations in $(1-\phi)/2$ around its minimum $(1-\phi_0)/2=0$, we can relate the ensemble average in the LHS of the above equation to the stochasticity level, which is an rms value, $S_2(t)={1\over 2}(1-\phi)_{rms}$. As expected, therefore, as the available turbulent kinetic energy at scale $l$, $v_{l,L}^2$, increases (decreases), the stochasticity level increases (decreases). Similarly, as magnetic field energy at scale $l$, i.e., $B_l^2$, increases (decreases), stochasticity level decreases (increases)\footnote{Note that our original definition of magnetic stochasticity does not rely on any ensemble averaging, therefore the expression given by the RHS of (\ref{stochasticity}) is not exactly the same as $S_2(t)$ but rather we assume that its behavior resembles magnetic stochasticity in the weak field regime.};

\begin{equation}\label{stochasticity}
S_2(t)|_{weak field}\rightarrow {1\over 2}\Big(1- {2\over 3}{B_lB_L\over v_{l,L}^2 } \Big)_{rms}.
\end{equation}

Thus as mean energy $\chi=B_l B_L$ increases (decreases), the stochasticity level decreases (increases). The mean cross energy density is defined as $E_2(t)={1\over 2}(B_lB_L)_{rms}$, hence in the weak field regime of $B_lB_L\lesssim v_{l,L}^2$, we expect as the stochasticity $S_2(t)$ increases (decreases), the mean energy $E_2(t)$ will decrease (increase); see Fig.(\ref{sub1}).

\begin{figure}[t]
 \begin{centering}
\includegraphics[scale=.5]{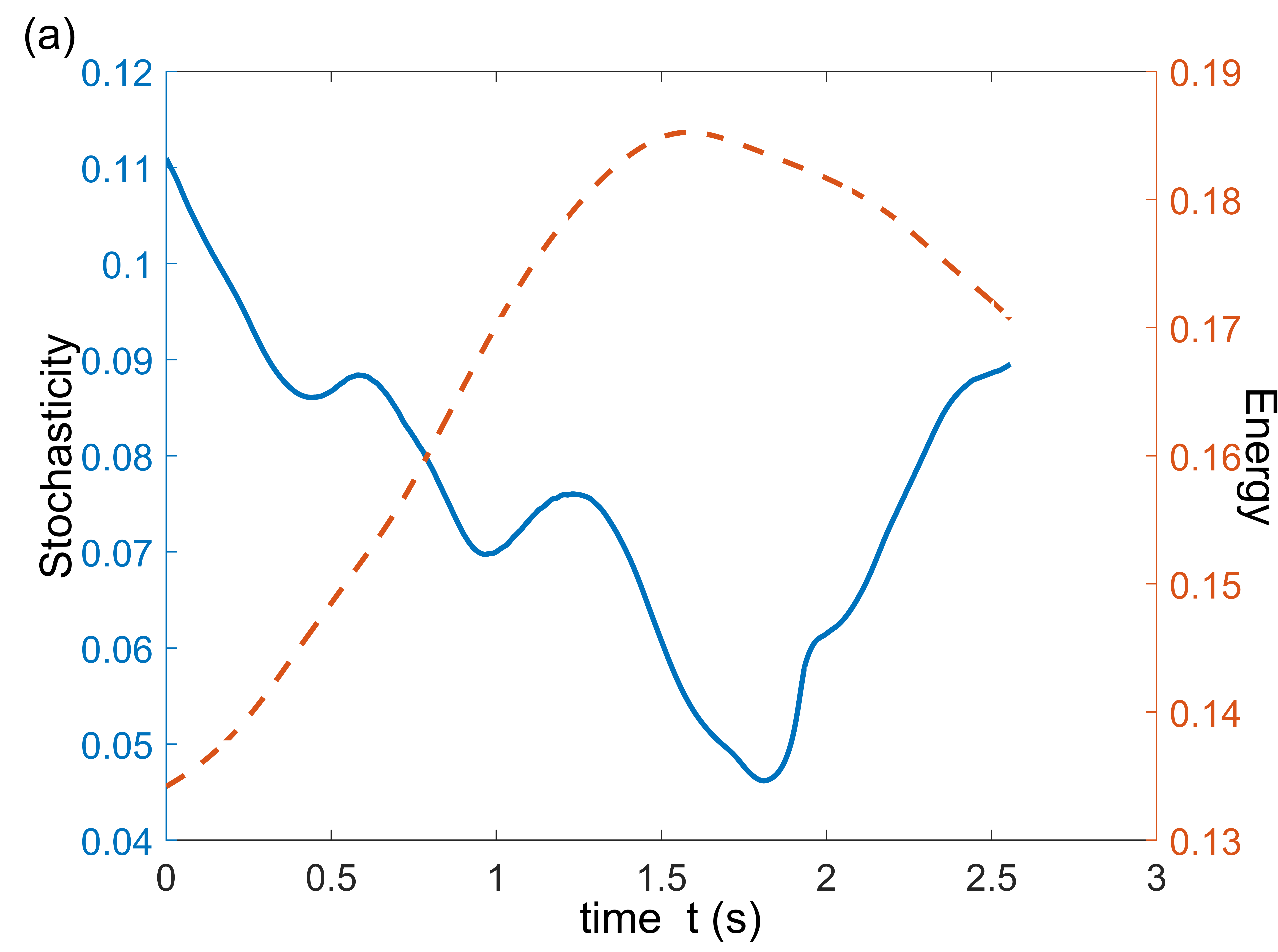}
\includegraphics[scale=.223]{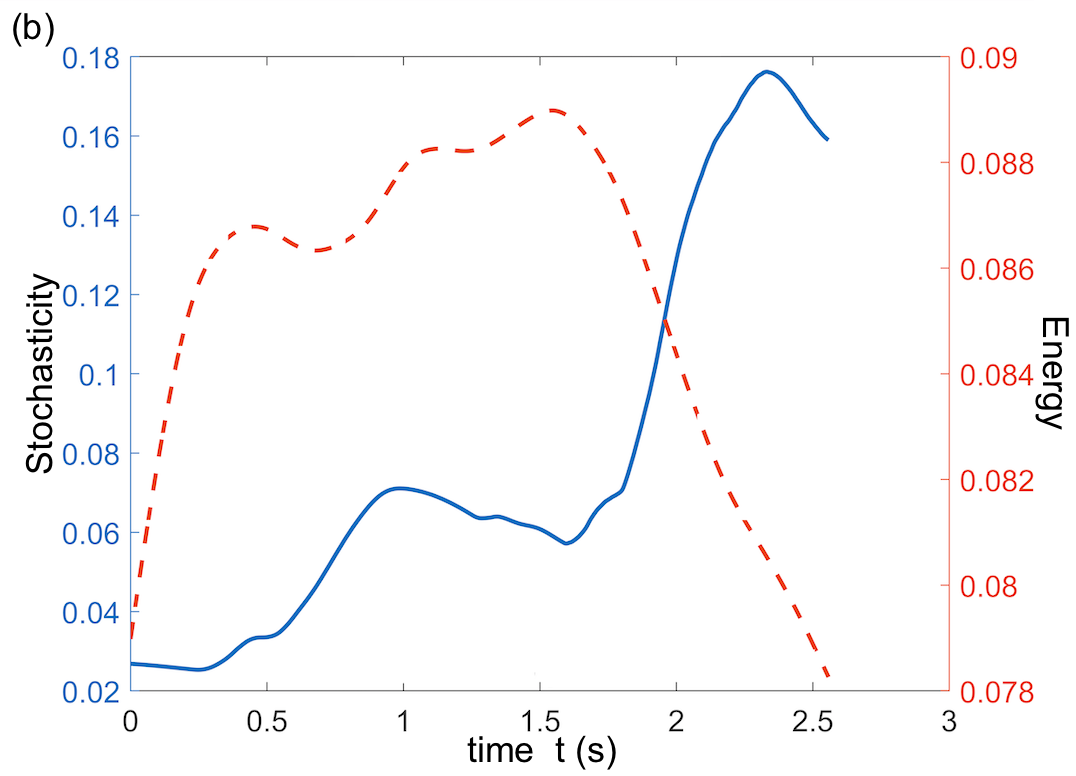}
\caption {\footnotesize {Stochasticity level $S_2(t)$ (blue, solid curve) and cross energy $E_2(t)$ (red, dashed curve) for two sub-volumes of the simulation box with small magnetic energy dissipation; $\Delta E_2\lesssim 0.05$ (a), $\Delta E_2\lesssim 0.01$ (b). These events might indicate field-fluid slippage and not reconnection which is accompanied with efficient magnetic energy dissipation by definition. This also suggests a method of categorizing different magnetic phenomena; see Table (\ref{table1}). }}\label{quietSV}
\end{centering}
\end{figure}

\subsection{Energy and Stochasticity Relaxation}

The above arguments imply that if we imagine a magnetized medium of scale $L$ as an ensemble of magnetized fluid parcels of scale $l\ll L$, similar to an ensemble of magnets embedded in the mean field generated by all neighbor magnets  (classical version of Ising model in mean field approximation), each parcel with average local field ${\bf B}_l$ will tend to align itself with the large scale field ${\bf B}_L$. This translates into the fact that locally the field tends to increase the scalar field $\phi=\hat{\bf B}_l.\hat{\bf B}_L$ or lower the stochasticity $S_2(t)=(1-\phi)_{rms}/2$. The magnetic field has a tendency to lower its stochasticity level similar to its tendency to lower its energy level. Because $\psi={1\over 2} {\bf B}_l.{\bf B}_L$ contains information about both the magnetic energy (through $\chi={1\over 2}B_l B_L$) and magnetic stochasticity (through $\phi=\hat{\bf B}_l.\hat{\bf B}_L$), mathematically, this translates into the problem of finding the extrema of $\psi={1\over 2}{\bf B}_l.{\bf B}_L$ instead of ${1\over 2}B^2$ as in Taylor relaxation. This can be done using the Lagrangian ${\cal L}=\psi({\bf x}, t)$. However, if the magnetic field $\bf B$ is to satisfy the induction equation then the coarse-grained field ${\bf B}_l$ will satisfy the coarse-grained induction equation, given by eq.(\ref{induction1}). If we define the scale-split magnetic helicity as 

$$H_{l, L}={\bf A}_l.{\bf B}_L,$$

it follows that the quantity 

\begin{equation}
H_{[l, L]}={1\over 2}\Big( {\bf A}_l.{\bf B}_L-{\bf A}_L.{\bf B}_l   \Big),
\end{equation}

is strictly conserved:

\begin{equation}
{\partial H_{[l, L]}\over \partial t} +\nabla.{\cal{J}}_{l, L}=0,
\end{equation}
with flux ${\cal{J}}_{l, L}={\bf A}_L\times({\bf R}_l+{\bf P}_l-{\bf u}_l\times{\bf B}_l)-{\bf A}_l\times({\bf R}_L+{\bf P}_L-{\bf u}_L\times{\bf B}_L)-{\Phi\!\!\!/}_L{\bf B}_L+{\Phi\!\!\!/}_l{\bf B}_l$ where $\bf A$ is the vector potential and ${\Phi\!\!\!/}$ is the scalar potential (not to be confused with $\phi={1\over 2}\hat{\bf B}_l.\hat{\bf B}_L$ or $\Phi={1\over 2}\hat{\bf u}_l.\hat{\bf u}_L$ extensively used in this paper). This constraint can be introduced to the Lagrangian using a Lagrange multiplier $\lambda$. We find

\begin{equation}
{\cal L}=\psi+\lambda H_{[l, L]}= {1\over 2}{\bf B}_l.{\bf B}_L+{\lambda\over 2}\Big( {\bf A}_l.{\bf B}_L-{\bf A}_L.{\bf B}_l   \Big).
\end{equation}

Variation with respect to ${\bf A}_l$ and  ${\bf A}_L$ yields respectively ${\bf B}_l\propto \nabla\times {\bf B}_l$ and ${\bf B}_L\propto \nabla\times {\bf B}_L$. This is a generalization of Taylor relaxation process: turbulent magnetic fields tend to lower both their stochasticity level and energy, on all scales, to reach a force-free state.

\begin{figure}[h]
 \begin{centering}
\includegraphics[scale=.527]{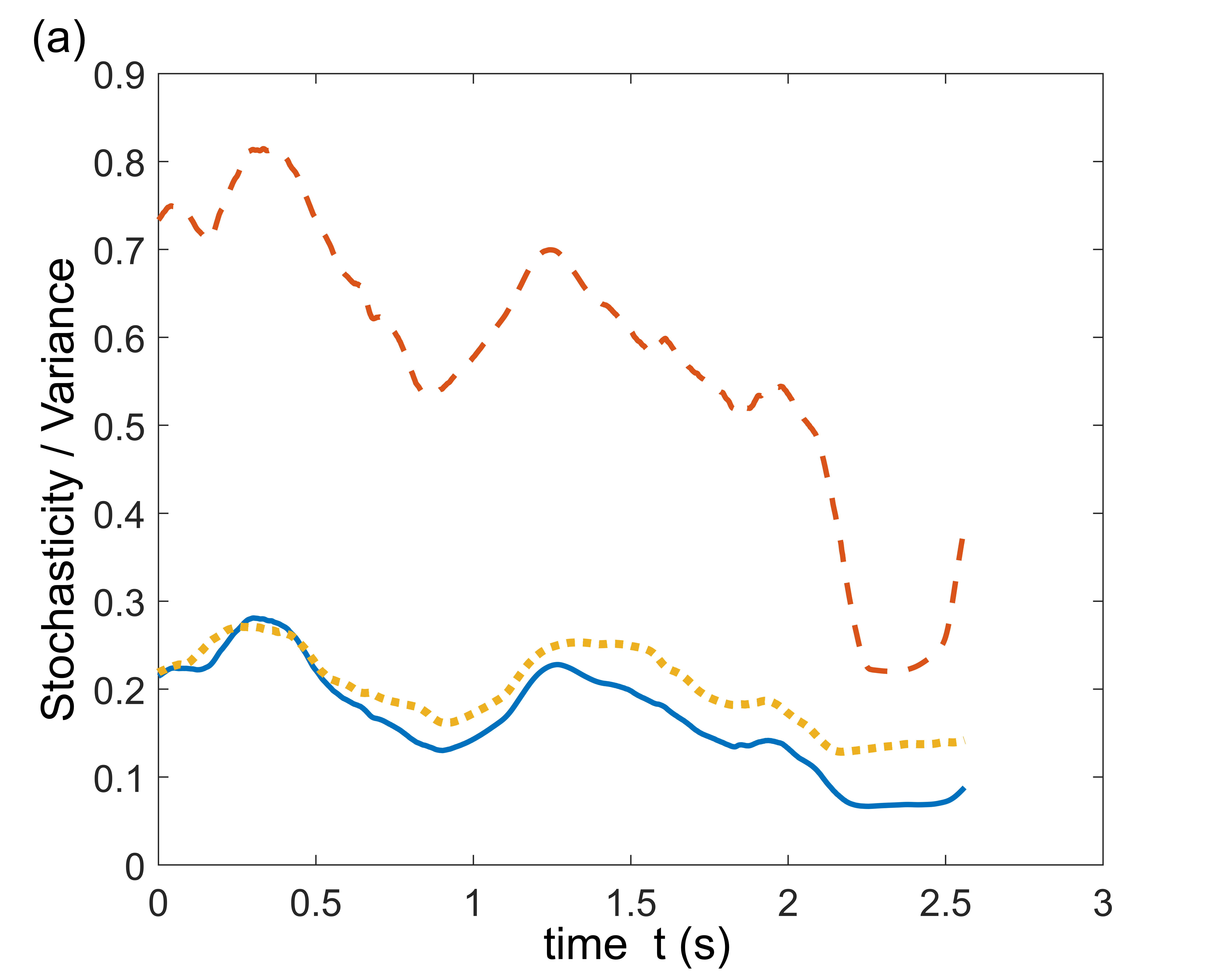}
\includegraphics[scale=.5]{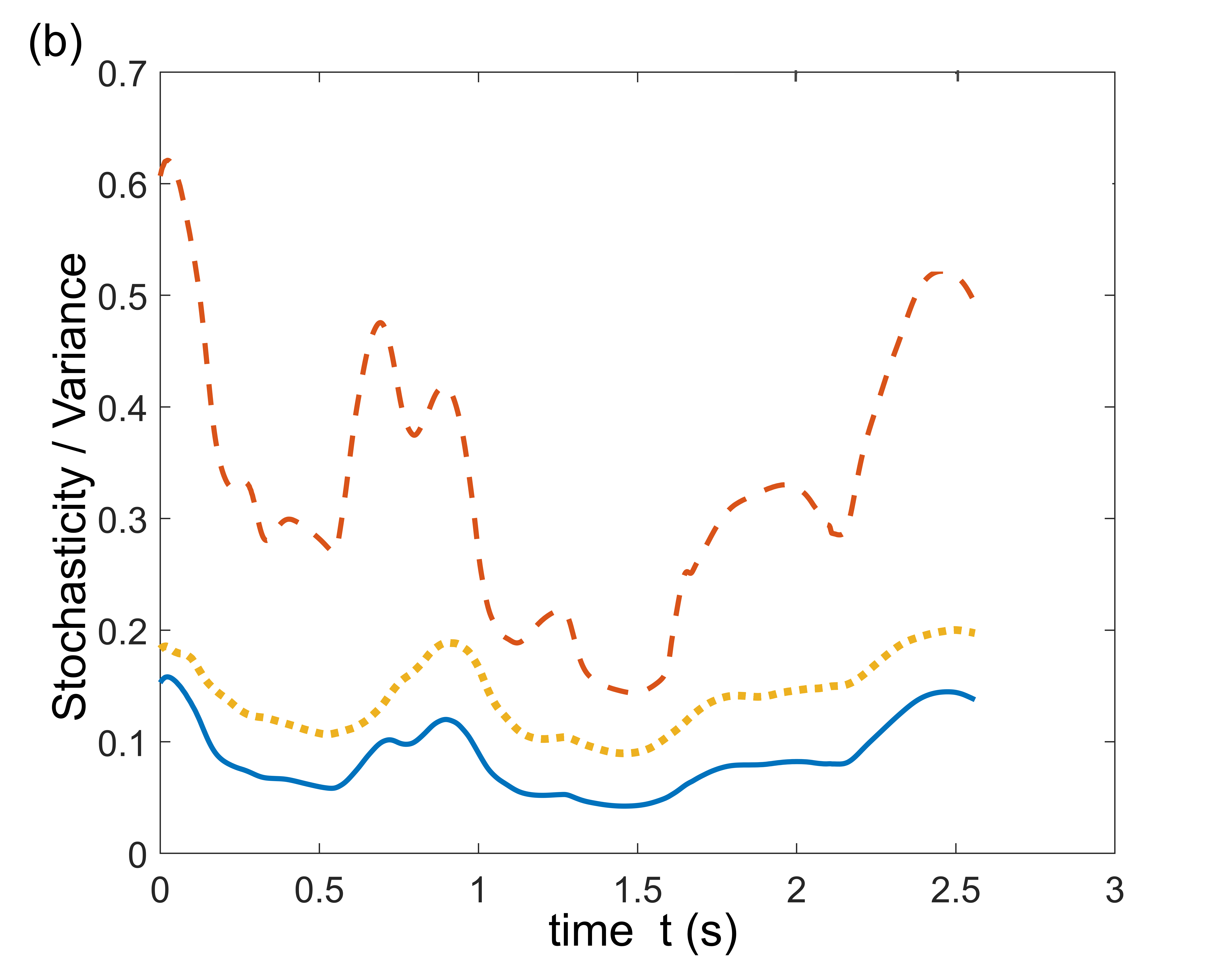}
\caption {\footnotesize {The stochasticity $S_2(t)=(1-\phi)_{rms}/2$ (blue, solid curve), the function $s_\phi=[(\phi-\phi_{rms})^2]_{rms}^{1/2}$ (red, dashed curve), and $s'_\phi=[(\phi^2)_{rms}-(\phi_{rms})^2]^{1/2}$ (yellow, dotted curve) in two different sub-volumes of the simulation box. Although different in terms of their absolute values, all three measures show a similar trend. These similarities are typical in different sub-volumes of the simulation box and indicate that the concept of magnetic stochasticity $S_2(t)$ can be used as a measure of standard deviation with the field's RMS value taken as the mean. }}\label{variance}
\end{centering}
\end{figure}

In passing, we also note that a more familiar way to define stochasticity may seem to be

\begin{equation}\label{StochasticityDef}
s_\phi^2=[(\phi-\phi_{rms})^2]_{rms},
\end{equation}
which is not incidentally equivalent to

\begin{equation}\label{StochasticityDef2}
s_\phi'^2=(\phi^2)_{rms}-(\phi_{rms})^2.
\end{equation}
These are similar to the definition of variance in probability theory and statistics except for rms averaging instead of taking the expectation value;

\begin{equation}\notag
\sigma_X^2=[\overline{(x-\overline x)^2}]=\overline {x^2}-\overline {x}^2,
\end{equation}
where $\overline x=E[X]$ is the expected value of random variable $X$. In our simulation, $\phi_{rms}\lesssim 1$, and a comparison of $s_\phi$ with $S_2(t)={1\over 2}(1-\phi)_{rms}$ in Fig.(\ref{variance}) shows that in fact these definitions have a very similar behavior. The other reason that we prefer the definition $S_2(t)={1\over 2}(1-\phi)_{rms}$, besides its simplicity, is that we are interested in measuring the deviation of $\phi=\hat{\bf B}_l.\hat{\bf B}_L$ from unity (which corresponds to a vanishing stochasticity) not its deviation from an average value $\phi_{rms}$.

\subsection{Slippage, Reconnection and Field Annihilation}\label{IIIC}

Let us consider the evolution of magnetic energy in terms of stochasticity level $S_2(t)$ and mean cross energy $E_2(t)$. It is easy to show that for $0<l\ll L$, ${B_L^2({\bf x}, t)}\leq {B_l^2({\bf x}, t)}\leq {B^2({\bf x}, t)}$ and therefore \footnote{Because for $l\ll L$, we have $G(r/L)\leq G(r/l)$ for $0\leq r <\infty$, because $G$ is by assumption a rapidly decaying, non-negative function, thus we find
 $$\int_V {d^3r\over L^3}G(r/L)|{\bf B}({\bf x+r})|\leq \int_V {d^3r\over l^3} G(r/l) |{\bf B}({\bf x+r})|,$$ 
 which means $B_L=|{\bf B}_L|\leq |{\bf B}_l|=B_l$. If we also note that the bare field is defined as $B=\lim_{l\rightarrow 0} B_l$, it follows that $B_L({\bf x}, t) \leq B_l({\bf x}, t) \leq B({\bf x}, t)$.}

\begin{equation}\notag
\chi({\bf x}, t)={|{\bf B}_l ({\bf x}, t)||{\bf B}_L({\bf x}, t)|\over 2}\leq {B^2({\bf x}, t)\over 2},
\end{equation}
which in turn leads to
\begin{equation}
E_2(t)={1\over 2}(B_l B_L)_{rms} \leq {1\over 2}(B^2)_{rms}.
\end{equation}

\begin{figure}[t]
 \begin{centering}
\includegraphics[scale=.23]{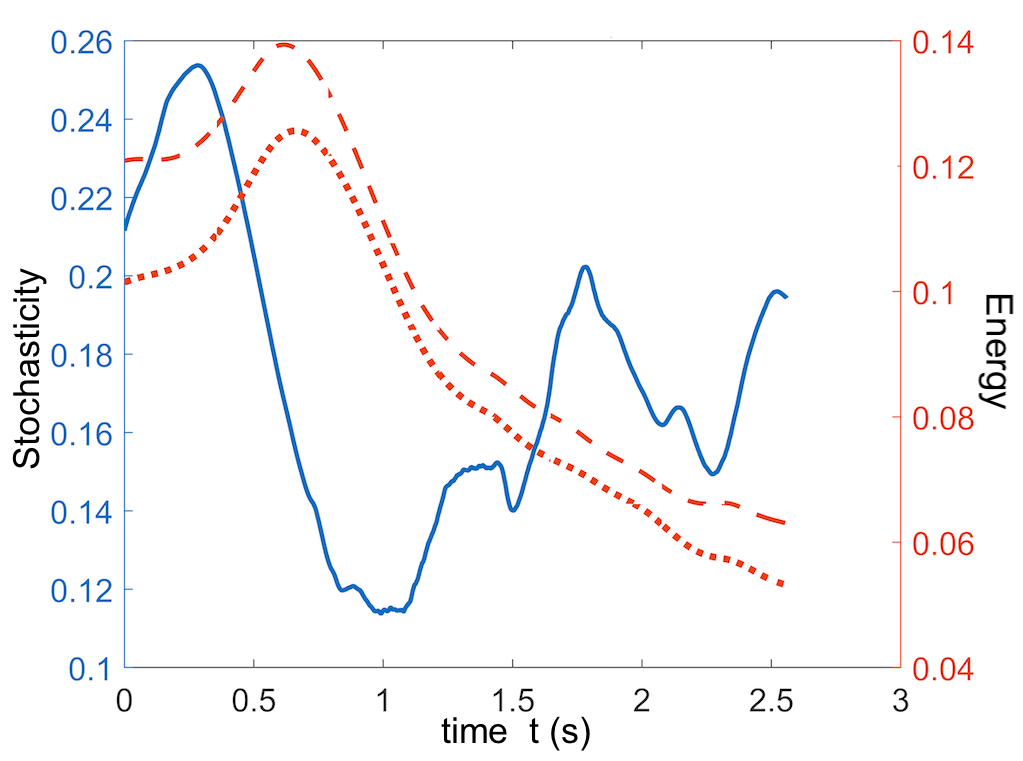}
\caption {\footnotesize {Typical behavior of magnetic stochasticity level $S_2(t)$ (blue, solid curve), cross energy $E_2(t)$(red, dotted curve) and RMS energy $(B^2/2)_{rms}$ (red, dashed curve). This plot also shows that the cross energy $E_2(t)$ evolves similar to the real magnetic energy density $(B^2/2){rms}$. In turns out that in fact the large scale, $(B_L^2/2)_{rms}$, as well as small scale $(B_l^2/2)_{rms}$ energy densities show similar trends. }}\label{measures}
\end{centering}
\end{figure}

In Fig.(\ref{measures}), we have plotted $E_2(t)$ and $(B^2/2)_{rms}$ in two sub-volumes of the simulation box. The top plot also shows few other measures of magnetic energy at two different scales. The time evolution of the cross energy is obviously very similar to that of real mean energy $(B^2/2)_{rms}$, although it is smaller numerically. Similar behavior is observed in other sub-volumes of the simulation box. This implies that an increasing (decreasing) stochasticity level $S_2(t)$ is accompanied with a decreasing (increasing) mean energy density $(B^2/2)_{rms}$ with the minima (maxima) of each one almost coincident with the maxima (minima) of the other one. 

If (i) local magnetic field reversals are ubiquitous in MHD turbulence \citep*{LazarianandVishniac1999} and (ii) magnetic reconnection occurs on all scales and is intimately related to field-fluid slippage (see \citep*{Eyink2015}), then the picture outlined above \citep*{JV2019} suggests that the maxima of stochasticity level $S_p(t)$ should approximately coincide with minima of mean cross energy density $E_p(t)$. A magnetic reconnection event in volume $V$ may be associated with

\begin{equation}\label{conditions2}
\Big\{T_2=D_2=0; \;\;\partial_t T_2 \leq 0,\;\&\;\partial_t D_2\geq 0\Big\}.
\end{equation}

We may also define the slippage intensity, for the time $\tau$ during which $S_p(t)$ changes considerably: 
\begin{eqnarray}\label{RecIntensity}
I_2(\tau)=\Big| [S_2(t_0+\tau)-S_2(t_0)]\Big|.
\end{eqnarray}

Note that generally field-fluid slippage may or may not be associated with magnetic null points. If it is, and the above conditions hold, magnetic field lines disconnect and reconnect again, therefore, close points on the field lines will not generally remain close to one another as the field lines disconnect. Hence magnetic reconnection is field-fluid slippage in which magnetic energy is reduced, magnetic connectivity breaks apart and topology changes. Topological deformation $T_p$ then also indicates topology change. 

The above arguments suggest also the following categorization for magnetic phenomena: (a) magnetic field annihilation, during which magnetic energy is dissipated with no significant change in magnetic topology, might correspond to considerable change in cross energy $\Delta E_2\gg 0$ but not in stochasticity $\Delta S_2\sim 0$; (b) during a field-fluid slippage, stochasticity changes significantly $\Delta S_2\gg 0$ but not cross energy $\Delta E_2\sim 0$; (c) during a global magnetic reconnection both stochasticity and cross energy change significantly $\Delta S_2\gg 0$, $\Delta E_2\gg 0$; and finally (d) for local, small scale reconnection events ubiquitous in MHD turbulence we expect both stochasticity and energy change to be small $\Delta S_2\sim 0$, $\Delta E_2\sim 0$. In all cases, we expect cross energy $E_2$ trace the rms magnetic energy density with a similar behavior and the relationship between $S_2$ and $E_2$ persists almost always; see Table.(\ref{table1}).

\begin{table}[t]
\caption{Hypothetical categorization of different magnetic phenomena based on the total variation of stochasticity and magnetic energy as a large or small fraction of unity.}
\centering
\begin{tabular}{c  c c c c}
\hline\hline
$Reconnection$  & $Slippage$ & $\;$ & $Annihilation$ & $Local \;Reversals$ \\ [0.6ex] 
\hline
large $\Delta S_2(t)$&large $\Delta S_2(t)$&\;&small $\Delta S_2(t)$&small $\Delta S_2(t)$\\
large $\Delta E_2(t)$ & small $\Delta E_2(t)$ &\;&large $\Delta E_2(t)$&small $\Delta E_2(t)$ \\ [1ex]
\hline
\end{tabular}
\label{table1}
\end{table}

\subsection{Topology Change and Reconnection}

The concepts of stochasticity level and cross energy can be applied to any other vector field including the velocity field ${\bf u}({\bf x}, t)$ in a turbulent fluid. In this case, we can define kinetic stochasticity $s_p(t)$, and kinetic cross energy $e_p(t)$ respectively as

\begin{equation}\label{formulae}
s_p(t)={1\over 2} ||1-\Phi({\bf x}, t) ||_p,
\end{equation}
and
\begin{equation}\label{formulae2U}
e_p(t)= ||{\cal{X}}({\bf x}, t) ||_p.
\end{equation}
Here, we have renormalized the velocity field ${\bf u}({\bf x}, t)$ at two scales $l$ and $L\gg l$ to define the scale-split kinetic energy density

\begin{equation}\label{scale-splitU}
\Psi_{l,L}({\bf{x}}, t)={1\over 2} \;{\bf{u}}_l({\bf{x}}, t){\bf{.u}}_L({\bf{x}}, t).
\end{equation}

\begin{figure}
 \begin{centering}
\includegraphics[scale=.53]{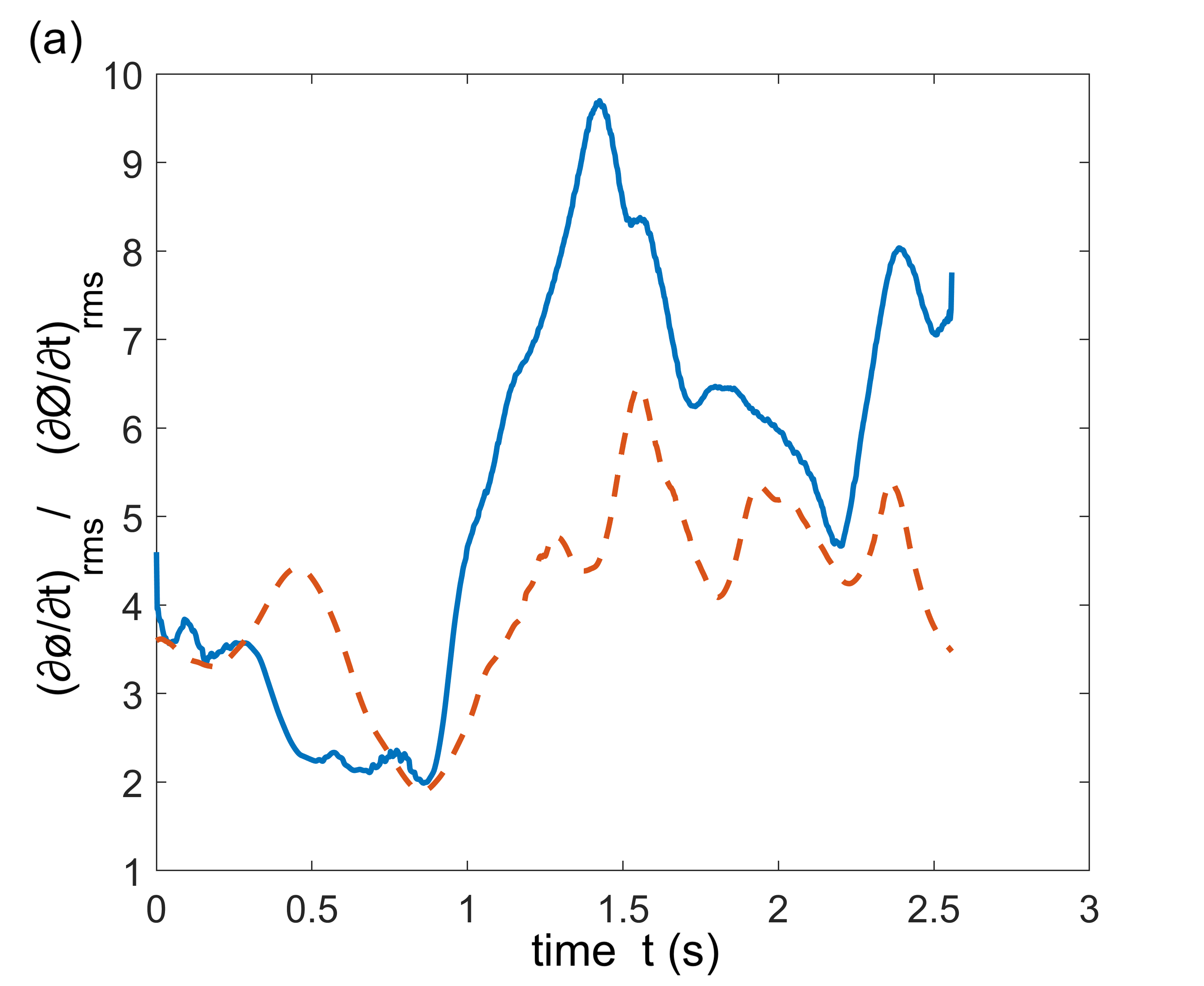}
\includegraphics[scale=.53]{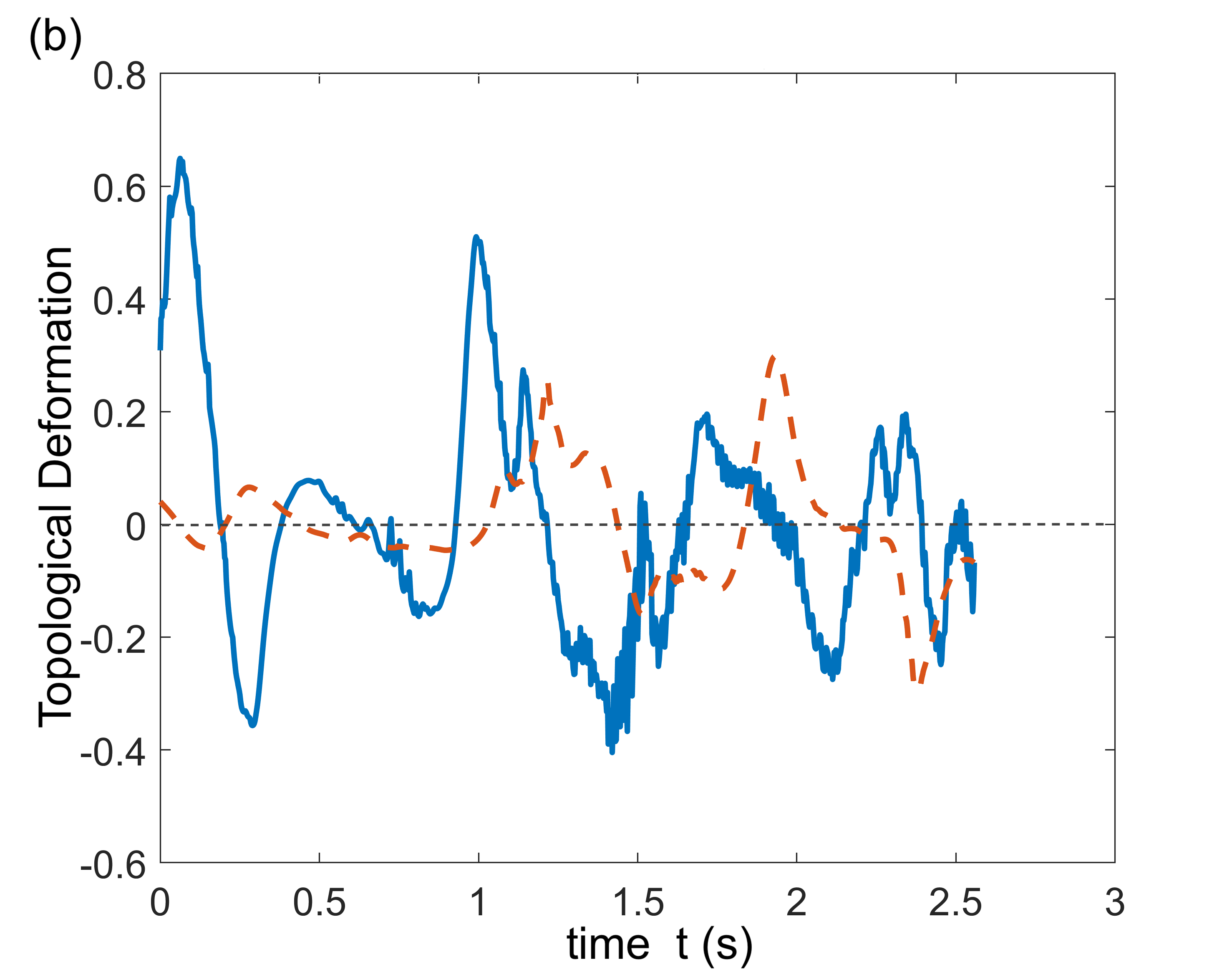}
\caption {\footnotesize {(a) The rms values of $\partial_t\phi$ (blue, solid curve) and its kinetic counterpart $\partial_t\Phi$ (red, dashed curve), where $\phi=\hat{\bf B}_l.\hat{\bf B}_L$ and $\Phi=\hat{\bf u}_l.\hat{\bf u}_L$ are magnetic and kinetic topology fields respectively. We have multiplied $\partial_t\Phi$ by a numerical factor of $\sim 4$ to make the comparison easier. Clearly, $(\partial_t\Phi)_{rms}$ is correlated with but falls behind $(\partial_t\phi)_{rms}$ with an almost constant time delay. This correlation translates into a correlation between magnetic and kinetic topological deformations defined as $T_2=\partial_t S_2$ and $\tau_2=\partial_t s_2$ since they are weighted volume-averages of $\partial_t\phi$ and $\partial_t\Phi$ respectively. (b) Magnetic (blue, solid curve) and kinetic (red, dashed curve) topological deformations, $T_2$ and $\tau_2$, in the same volume. When $T_2=\partial_t S_2=0\;\&\; \partial_t T_2=\partial_t^2 S_2<0$, shown by black dots, the magnetic stochasticity reaches a maximum and magnetic reconnection peaks. As magnetic stochasticity starts to decrease, we have $T_2<0$ and the reconnecting field lines push the fluid and increase the kinetic stochasticity; $\tau_2=\partial_t s_2>0$.  }}\label{dphiT1}
\end{centering}
\end{figure}

\begin{figure}
 \begin{centering}
\includegraphics[scale=.48]{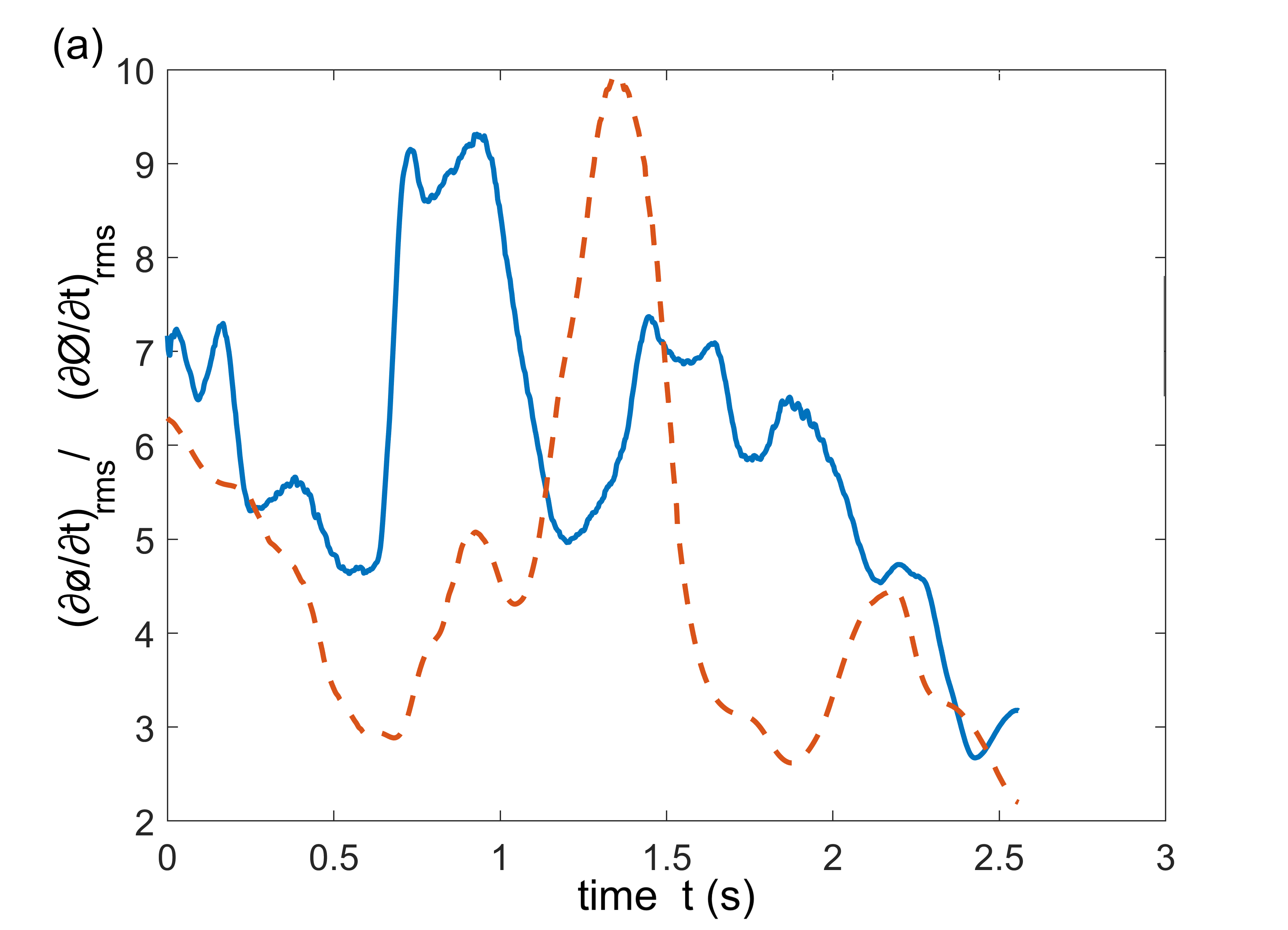}
\includegraphics[scale=.565]{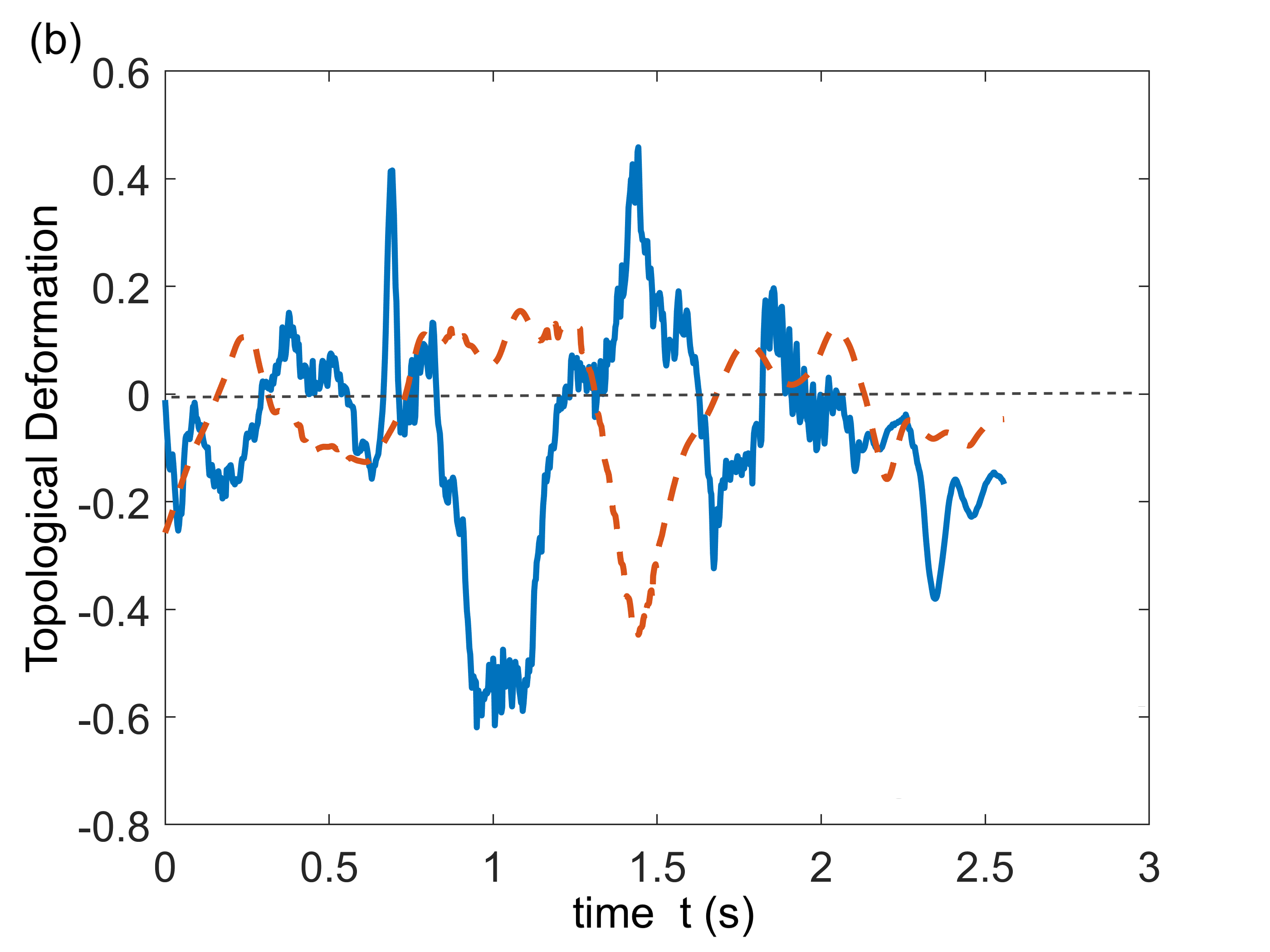}
\caption {\footnotesize {Same as Fig.(\ref{dphiT1}) but for a different sub-volume of the simulation box. (a) Typically, we see a phase shift or time delay between $(\partial_t\phi)_{rms}$ (blue, solid curve) and $(\partial_t\Phi)_{rms}$ (red, dashed curve), which may indicate the interaction of reconnecting magnetic field and the fluid. This effect is easier to interpret in terms of the magnetic and kinetic topology deformations, i.e., $T_2$ and $\tau_2$. (b) As magnetic stochasticity $S_2$ (blue, solid curve) increases, i.e., $T_2=\partial_t S_2>0$, by the tangling effect of the turbulence, it reaches a maximum where $T_2=0\;\&\;\partial_t T_2<0$. The magnetic field starts to reconnect and reduce its stochasticity, so $T_2<0$, which leads to the ejection of the fluid of the region. This in turn increases the kinetic stochasticity (red, dashed curve) of the turbulent velocity field $\bf u$, hence $\tau_2=\partial_t s_2>0$.}}\label{dphiT2}
\end{centering}
\end{figure}

Similar to the magnetic scale-split energy $\psi$, the kinetic scale-split energy $\Psi$ too can be divided into two scalar fields; the kinetic topology field,
 
 \begin{equation}\label{phichi1U}
\Phi_{l,L} ({\bf{x}}, t)=\begin{cases}
\hat{\bf{u}}_l({\bf{x}}, t).\hat{{\bf{u}}}_L({\bf{x}}, t) \;\;\;\;\;\;u_L\neq 0\;\&\;u_l\neq0,\\
0\;\;\;\;\;\;\;\;\;\;\;\;\;\;\;\;\;\;\;\;\;\;\;\;\;\;\;\;\;\;\;\;\;otherwise.
\end{cases}
\end{equation}
and the kinetic energy field,

\begin{equation}\label{phichi2U}
{\cal X}_{l,L}({\bf{x}}, t)={1\over 2} u_l ({\bf{x}}, t) u_L({\bf{x}}, t).
\end{equation}

As before, we may take $p=2$ for simplicity, in which case the kinetic stochasticity level $s_2$, kinetic topological deformation $\tau_2=\partial_t s_2$, kinetic cross energy density $e_2(t)$, and kinetic energy dissipation $d_2=\partial_t e_2$ are given by

\begin{equation}\label{formulaeU}
s_2(t)={1\over 2} (1-\Phi  )_{rms},
\end{equation}

\begin{equation}
\tau_2(t)=  {1\over 4 s_2(t)} \int_V \;(\Phi-1){\partial \Phi\over \partial t}\; {d^3x\over V},
\end{equation}

\begin{equation}
e_2(t)={\cal X}_{rms},
\end{equation}

and
\begin{equation}
d_2(t)={1\over  e_2(t)}\int_V {\cal X} \partial_t {\cal X}{d^3x\over V}.
\end{equation}

It follows that 

\begin{eqnarray}\notag
\tau_2(t)=&&{1\over 4 s_2(t)}  \int_V   \Big[\hat{\bf u}_l.\hat{\bf u}_L-1 \Big] \; \Big[  \hat{\bf{u}}_L. \Big({\partial_t {\bf u}_l \over u_l} \Big)_{\perp {\bf u}_l }  \\\label{T1U}
&& + \hat{\bf{u}}_l. \Big({\partial_t {\bf u}_L\over u_L}  \Big)_{\perp{\bf u}_L }          \Big]  \;{d^3x\over V}.
\end{eqnarray}

Here, $(\;)_{\perp{\bf{u}}}$ represents the perpendicular component with respect to $\bf{u}$. In a similar way, we find

\begin{eqnarray}\notag
d_2(t)&=&{1\over e_2(t)}\int_V  \Big({u_l u_L\over 2}\Big)^2\\\label{Edissipation300U}
&\times&  \Big[{\partial_t (u_L^2/2)\over u_L^2}+{\partial_t (u_l^2/2)\over u_l^2}    \Big]{d^3x\over V}.
\end{eqnarray}

The time evolution of the topology field $\phi({\bf x}, t)=\hat{\bf B}_l.\hat{\bf B}_L$ gives us important information about the changes in the field configuration. More precisely, the time derivative of $\phi({\bf x}, t)$ corresponds to the local topological deformations (or changes) at point ${\bf x}$ at time $t$. The top panel in Fig.(\ref{dphiT1}) shows the rms value of the time derivative of magnetic topology field, i.e., $(\partial\phi/\partial t)_{rms}$, as well as its kinetic counterpart; $(\partial \Phi/\partial t)_{rms}$. There is a clear correlation between the time derivatives of the magnetic and kinetic topology fields but, more importantly, there is some delay, or phase shift, between the two functions: the kinetic topology change seems to lag behind the magnetic topology change. See also the top panel in Fig.(\ref{dphiT2}).

The topological deformation $T_2(t)$, given by eq.(\ref{Tdeform}), is a weighted average of $\partial_t \phi$. On the other hand, it is the time derivative of the stochasticity level $S_2(t)$ which is in turn related to the cross energy $E_2(t)$, as discussed in the previous section. The bottom panel in Fig.(\ref{dphiT1}) shows a typical graph of magnetic topological deformation function $T_2$ with its kinetic counterpart $\tau_2(t)=\partial_t s_2(t)$.

Turbulence tends to increase the magnetic stochasticity by tangling field lines. The increasing stochasticity reaches a maximum level, when $T_2=\partial_t S_2=0$ and $\partial_t T_2=\partial_t^2 S_2<0$. As magnetic field reconnects to relax to a lower stochasticity, the topological deformation becomes negative $T_2<0$. Reconnecting field lines push the fluid and increase the kinetic stochasticity $s_2$, i.e., $\tau_2=\partial_t s_2>0$: see also the bottom panel in Fig.(\ref{dphiT2}).

The local changes in $T_2$, corresponding to local field reversals, may cancel one another out when calculated in a large volume. In other words, since $T_2$ is a weighted integral of $\partial_t\phi$ over an arbitrary volume $V=L^3$, the out of phase topological changes in different regions of scale $l<L$ inside the volume $V=L^3$ may cancel out when summed over. In that case, we have local reconnection events occurring locally at small scales of order $l$. If, on the other hand, the topological changes ongoing in different regions inside $V$ are in phase, they would give rise to an appreciable total $T_2$ in the whole volume $V=L^3$. This case may correspond to a global reconnection event at scale $L$.

\section{Summary and Conclusions}\label{sSummary}

In this paper, we have numerically tested the theoretical predictions made in the previous work on magnetic stochasticity \citep*{JV2019} and extended its arguments by including the kinetic stochasticity associated with the velocity field. One theoretical prediction in this formalism is that magnetic field slippage through the fluid, as well as magnetic reconnection which is also related to field-fluid slippage, should be accompanied with increasing stochasticity level and decreasing magnetic energy followed, after reaching their extrema, by a decreasing stochasticity level and increasing mean magnetic energy. This formalism is based on a simple scalar field, the scale-split magnetic energy density; $\psi={1\over 2} {\bf B}_l.{\bf B}_L$. We have also presented a simple toy model in order to illustrate how turbulence can in principle increase the randomness in the magnetic field in the weak-field regime. In this model, the scalar field $\psi$ appears in a partition function as an interaction energy term. Nevertheless, despite its usefulness in illustration of the interaction between MHD turbulence and magnetic fields, this toy model should not be taken too literally.

In order to test the above theoretical arguments, we have used the data from a homogeneous, incompressible MHD simulation stored online (\citep*{JHTDB};\citep*{JHTB1};\citep*{JHTB2}). The predicted pattern is observed in different sub-volumes of the simulation box implying that field-fluid slippage and local reconnections are an inseparable aspect of MHD turbulence. The statistical relationship between magnetic stochasticity and energy, described above, persists almost for all cases. In addition, the relationship between magnetic and kinetic topological deformations, defined as the time derivative of magnetic and kinetic stochasticities respectively, is observed to be in good agreement with the theory. A fast decrease in magnetic stochasticity after reaching its maximum value is almost always followed by a rapid increase in the kinetic stochasticity. This may indicate local reconnection events in which an initially tangled field (large stochasticity) decreases its stochasticity by reconnection, which in turn pushes the fluid and increases its kinetic stochasticity.

Our numerical findings in this paper in general agree with the theoretical predictions of Jafari and Vishniac \cite{JV2019}, made by applying their general formulation of stochastic vector fields to turbulent magnetic fields. Thus this formalism may be an interesting and fruitful way of studying turbulent magnetic fields. However, our results do not prove that magnetic stochasticity and magnetic energy in MHD turbulence always evolve consistently following a simple pattern, which, if true, can be useful in the study of magnetic reconnection and other magnetic phenomena such as magnetic dynamo. More numerical studies are required in order to decide whether the theoretical formulation of stochasticity and energy presented in \citep*{JV2019} and this paper is indeed useful in such problems. Finally, we should also mention an exception observed in our study of the relationship between $S_2(t)$ and $E_2(t)$. We have looked at more than $20$ randomly selected sub-volumes of the simulation box with different sizes, in all of which the predicted pattern is observed although in one small sub-volume, this relationship is not so obvious. This might be due to intermittency or other non-linear effects. We interpret this as a small deviation from a general pattern in a statistical sense, however, further studies might indicate otherwise pointing to something more serious.

The most important implications of this paper may be briefed as follows:

1. Turbulence introduces randomness to magnetic fields. Magnetic stochasticity can be quantified and related to magnetic topology and energy using the scalar field $\psi={1\over 2} {\bf B}_l.{\bf B}_L=\phi \chi$. In particular, the component $\phi=\hat{\bf B}_l.\hat{\bf B}_L$ is related to magnetic topology and stochasticity while $\chi={1\over 2} B_lB_L$ introduces a measure of magnetic energy. Magnetic stochasticity, or spatial complexity, is defined then as $S_2(t)=(1-\phi)_{rms}/2$ or in general as an ${\cal L}_p$-norm; $S_p(t)=||1-\phi||_p/2$.

2. Magnetic stochasticity and energy evolve accordingly following a simple pattern: increasing (decreasing) stochasticity almost always coincides with decreasing (increasing) magnetic energy. This relationship arises as a consequence of (a) the persistent slippage of magnetic field through the fluid and (b) small scale, local magnetic reconnection events in MHD turbulence. These two phenomena, i.e., field-fluid slippage and local reconnections, are in fact related: the former has been formulated by Eyink \citep*{Eyink2015} and shown to be intimately connected to magnetic reconnection while the latter, i.e., the phenomenon of local small-scale reconnections in MHD turbulence, has been formulated as the base of stochastic reconnection model by Lazarian and Vishniac \citep*{LazarianandVishniac1999}. 

3. Magnetic reconnection seems to correspond to the simultaneous extrema of magnetic stochasticity $S_2(t)$ and magnetic energy $(B^2/2)_{rms}$ (whose evolution typically resembles that of $\chi=B_lB_L/2$ with $l\ll L$ and $L$ both chosen in the inertial range). This also suggests a hypothetical categorization of different magnetic phenomena, such as magnetic energy dissipation, reconnection and field-fluid slippage, in terms of the changes in magnetic stochasticity and energy. Hence, for example, large variations in both stochasticity and energy will imply reconnection whereas small variations in energy accompanied with large changes in stochasticity may imply field-fluid slippage. Magnetic field annihilation may also correspond to large decreases in energy but negligible changes in stochasticity. This categorization remains hypothetical in our work and requires further, more detailed numerical studies.

4. The above result suggests a statistical approach to the kinematics of magnetic reconnection in terms of the magnetic and kinetic topological deformations, respectively, defined as $T_2=\partial_t S_2$ and $\tau_2=\partial_t s_2$. Magnetic field is stochastically frozen into the fluid (\cite{JV2019}; \cite{Eyink2011}), hence, turbulence will tend in general to increase magnetic stochasticity by tangling the field. Since magnetic field resists bending and tangling by the turbulence, magnetic stochasticity cannot increase indefinitely and instead it reaches a maximum level, corresponding to $T_2=\partial_t S_2=0$ and $\partial_t T_2=\partial_t^2 S_2<0$. Magnetic reconnection can reduce the stochasticity level and let the field relax to a lower state. Decreasing stochasticity, in turn, means a negative topological deformation; $T_2<0$. The reconnecting field pushes the fluid and increases the kinetic stochasticity $s_2$, i.e., $\tau_2=\partial_t s_2>0$. Combined with the relationship between magnetic stochasticity and cross energy discussed before, this provides a statistical representation of magnetic reconnection in terms of the magnetic and kinetic stochasticities, their time derivatives and also corresponding magnetic and kinetic cross energies. Overall, our numerical results are in good agreement with this picture. 

5. Stochasticity $S_2(t)$ and cross energy $E_2(t)=(B_lB_L/2)_{rms}$, used to study reconnection/slippage on arbitrary scales $l<L$, are scale dependent functions in the turbulence inertial range. Application of the Renormalization Group (RG) invariance then leads to the conclusion that magnetic reconnection and field-fluid slippage in fact occur on a wide range of scales in the inertial range of MHD turbulence. In particular, it asserts that magnetic reconnection is not confined into small, dissipative regions. These provide, of course, another confirmation of the well-known fact that turbulent reconnection is a multi-scale phenomenon which at larger scales becomes totally independent of micro-scale effects like resistivity.

6) Similar to the scale-split energy $\psi={\bf B}_l.{\bf B}_L/2$, one can also define a scale-split magnetic helicity as $H_{l, L}={\bf A}_l.{\bf B}_L$ and the corresponding quantity 

$$H_{[l, L]}={1\over 2}\Big( {\bf A}_l.{\bf B}_L-{\bf A}_L.{\bf B}_l   \Big).$$
This quantity satisfies a continuity-like equation which can be used to formulate the stochasticity and energy relaxation of turbulent magnetic fields as a generalization of Taylor relaxation process.

\bibliographystyle{apsrev4-2}
\bibliography{StatisticalFinal2}

\end{document}